\documentclass[aps,pra,amsmath,amssymb,floatfix,twocolumn,amsmath,superscriptaddress,twocolumn,nofootinbib,tighten,letterpaper]{revtex4-2}
\usepackage{multirow}
\usepackage{bbold}

\usepackage{color}
\usepackage{mathrsfs}
\usepackage{hyperref}
\usepackage[normalem]{ulem}
\usepackage{bm}

\usepackage[caption=false]{subfig} % subfloat environment

\DeclareMathOperator{\sech}{sech}

\usepackage{amsfonts, relsize, color}
\usepackage{graphics}
\usepackage{graphicx}

\usepackage[table]{xcolor}

\usepackage{hyperref}
\usepackage{color}
\usepackage{comment}

\definecolor{RowColor}{rgb}{0.92,0.92,1.0}

\renewcommand\vec[1]{\ensuremath\boldsymbol{#1}} % bold font for vectors

\begin{document}
\title{Superconductivity in doped planar Dirac insulators: A renormalization group study}

\author{Sk Asrap Murshed}
\affiliation{Department of Physics, Lehigh University, Bethlehem, Pennsylvania, 18015, USA}

\author{Sanjib Kumar Das}
\affiliation{Department of Physics, Lehigh University, Bethlehem, Pennsylvania, 18015, USA}

\author{Bitan Roy}
\affiliation{Department of Physics, Lehigh University, Bethlehem, Pennsylvania, 18015, USA}

\date{\today}
\begin{abstract}
From a leading-order unbiased renormalization group analysis we here showcase the emergence of superconductivity (including the topological ones) from purely repulsive electron-electron interactions in two-dimensional doped Dirac insulators, featuring a Fermi surface. Otherwise a simply connected Fermi surface becomes annular deep inside the topological regime. In the absence of chemical doping, such systems describe quantum anomalous or spin Hall and normal insulators. By considering all symmetry allowed repulsive local four-fermion interactions, we show that the nature of the resulting superconducting states at low temperature follows certain Clifford algebraic selection rules, irrespective of the underlying Fermi surface topology. Within the framework of a microscopic Hubbard model, on-site repulsion among fermions with opposite orbitals (spin projections) typically favors odd-parity topological $p$-wave (conventional even-parity $s$-wave) pairing. Theoretically predicted superconductivity can in principle be observed in experiments once the promising candidate materials for quantum anomalous and spin Hall insulators are doped to foster Fermi surfaces, realizable in quantum materials and on optical lattices of cold atoms.   
\end{abstract}

\maketitle

%%%%%%%%%%%%%%%%%%%%%%%%%%%%%%%%%%%%%%%%%%%%%%%%%%%%%%%%%%%%%%%%%%%%%%%
%%%%%%%%%%%%%%%%%%%%%%%%%%%%%%%%%%%%%%%%%%%%%%%%%%%%%%%%%%%%%%%%%%%%%%%
%%%%%%%%%%%%%%%%%%%%%%%%%%% INTRODUCTION %%%%%%%%%%%%%%%%%%%%%%%%%%%%%%
%%%%%%%%%%%%%%%%%%%%%%%%%%%%%%%%%%%%%%%%%%%%%%%%%%%%%%%%%%%%%%%%%%%%%%%
%%%%%%%%%%%%%%%%%%%%%%%%%%%%%%%%%%%%%%%%%%%%%%%%%%%%%%%%%%%%%%%%%%%%%%%

\section{Introduction}

The massive Dirac Hamiltonian provides a universal description of topological insulators and superconductors belonging to any Altland-Zirnbauer symmetry class in any dimension. In this formulation the Dirac mass manifests a band inversion (changing its sign) at a finite momentum. Topologically trivial insulators and superconductors are also described by the massive Dirac Hamiltonian in which, however, the Dirac mass does not change its sign at any momentum~\cite{TITSC:1, TITSC:2, TITSC:3, TITSC:4, TITSC:5, TITSC:6, TITSC:7, TITSC:8,TITSC:9, TITSC:10, TITSC:11, TITSC:12, TITSC:13}. And the universality class of the quantum phase transition between topological and normal insulating states (electrical or thermal) is determined by a collection of massless Dirac fermions, characterized by the dynamic scaling exponent $z=1$ and the correlation length exponent $\nu=1$~\cite{TQPT:1, TQPT:2, TQPT:3}.

Dirac materials also constitute a suitable platform to harness topological superconductors (besides the featureless even-parity $s$-wave pairings). Namely, local or momentum-independent odd-parity superconducting Dirac masses, for which the corresponding operators fully anticommute with the Dirac Hamiltonian, inherit topology from the normal state and give rise to $p$-wave pairings~\cite{TSClocal:1, TSClocal:2, TSClocal:3, TSClocal:4, TSClocal:5, TSClocal:6, TSClocal:7}. When such pairings occur around the underlying Fermi surface (realized by chemically doping a Dirac insulator) it gets fully gapped. At the same time, the vector order parameter associated with the $p$-wave pairing nontrivially winds around the Fermi surface, yielding topological superconductivity.

\begin{figure}[t!]
\includegraphics[width=1.00\linewidth]{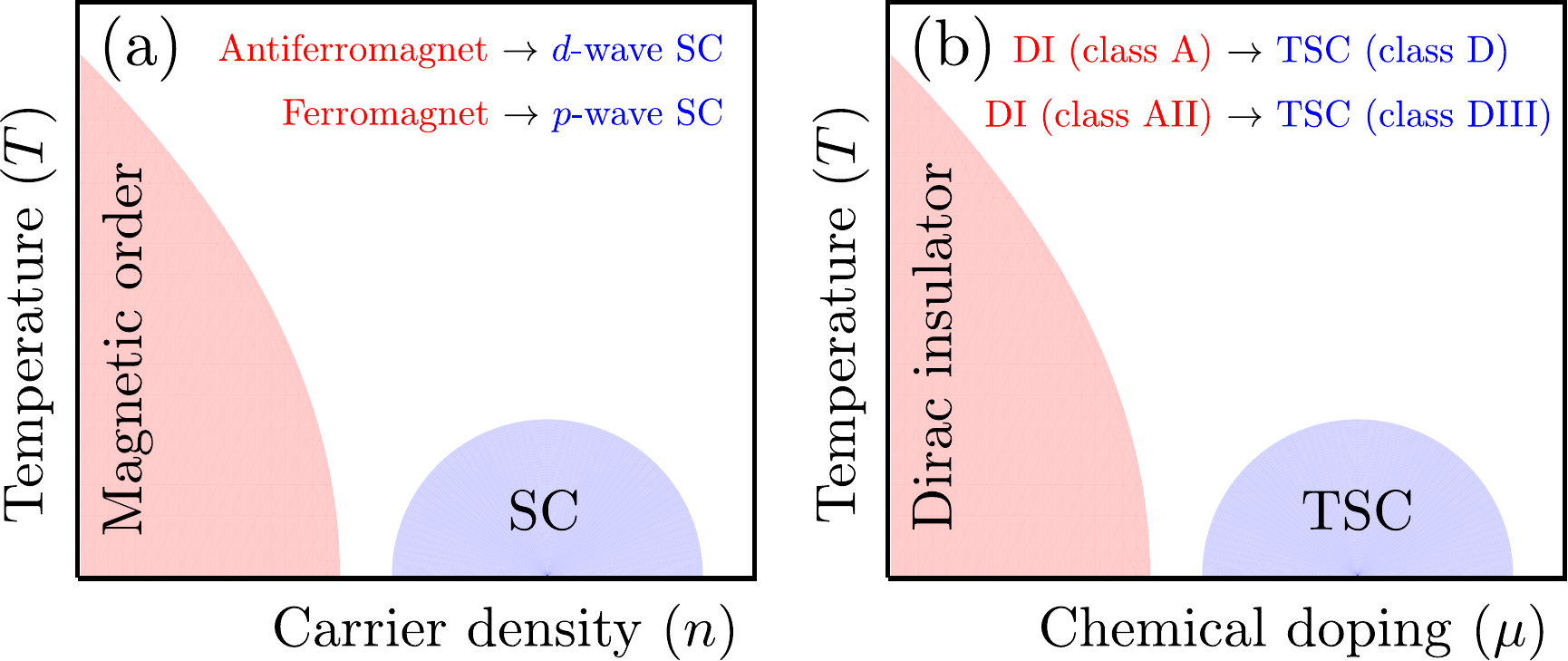}
\caption{(a) A schematic phenomenological phase diagram of strongly correlated doped magnetic materials, fostering a magnetic ground state at and near half-filling and a superconductor (SC) away from it, achieved by increasing the carrier density ($n$) in the system. Typically, a $d$-wave ($p$-wave) superconductor is found when the magnetic ground state is antiferromagnet (ferromagnet) that can be theoretically demonstrated from the conventional Hubbard model with on-site repulsion. (b) A proposed schematic phase diagram for interacting doped Dirac insulators (DIs) that can potentially harbor topological superconductors (TSCs) when the normal state is doped to accommodate a Fermi surface. Here $\mu$ corresponds to the chemical potential or the gate voltage. Theoretical study, presented in this work, suggests that an orbital Hubbard model with repulsive interactions among fermions with opposite orbitals or parity eigenvalues give rise to TSC belonging to class D (class DIII) when the parent DI belongs to class A (class AII) in two dimensions. We note that such an on-site orbital Hubbard repulsion in class AII Dirac system also yields a class DIII pairing in three dimensions~\cite{szaboroy:3DTI}. Although it is conceivable to find a class C topological nematic pairing upon doping an interacting planar class AII Dirac insulator, it goes beyond the realm of the Hubbard model. In (a), the magnetic order disappears typically via a true phase transition at finite temperature. In (b), the insulating behavior disappears as a crossover phenomena at finite temperature when it becomes comparable to the zero-temperature gap.           
}~\label{fig:philosophy}
\end{figure}

Fascinatingly, in this work we show that repulsive Hubbardlike local electron-electron interactions can conduce such topological pairing in doped Dirac insulators. In turn, this observation allows us to speculate a comparison between the global phase diagrams of strongly correlated doped magnetic materials and doped (topological) Dirac materials, shown in Fig.~\ref{fig:philosophy}, both fostering superconductivity at low temperatures. Therefore, a predictive confluence of electronic interactions and emergent topology can be established by concentrating on interacting doped Dirac insulators, from various Altland-Zirnbauer symmetry classes. So, we initiate the discussion with an overview on the Dirac theory.          

\subsection{Dirac Hamiltonian and Fermi surface}

The effective single-particle Hamiltonian describing a collection of noninteracting massive Dirac fermions in any spatial dimension ($d$) in the presence of a finite chemical potential or doping ($\mu$) takes the universal form 
\begin{equation}~\label{eq:DiracUniversal}
\hat{h}_{\rm Dir}(\vec{k})= \sum^{d}_{j=1} v_j \Gamma_j k_j + \left( m + \sum^{d}_{j=1} b_j k^2_j \right) \Gamma_{d+1} -\mu
\end{equation}
in the low-energy and long wavelength limit~\cite{Dirac:1, Dirac:2, Dirac:3}. Here, $v_i$ ($k_i$) is the Fermi velocity (momentum) along the $i$th coordinate. Throughout we assume a spatial rotational symmetry in the system, and thus set $v_j=v$ (say) and $b_j=b$ (say) for $j=1, \cdots, d$. The first term in Eq.~\eqref{eq:DiracUniversal} describes the Dirac kinetic energy that is linear in momentum, yielding $z=1$. In the second term, $m$ and $b k^2$ represent constant and momentum-dependent Wilson-Dirac masses. For simplicity, we ignore any particle-hole asymmetry in the $\mu=0$ limit. Mutually anticommuting Hermitian $\Gamma$ matrices satisfy the Clifford algebra $\{ \Gamma_j, \Gamma_k \}=2 \delta_{jk}$ for $j,k=1, \cdots, d+1$, where $\delta_{jk}$ is the Kronecker delta function. The dimensionality and representation of the $\Gamma$ matrices and the associated Dirac spinor depend on the dimensionality of the system and its symmetry class. Irrespective of these details, when $b>0$, which we assume throughout this work for concreteness, the above Hamiltonian describes a topological insulator for $m<0$ and a normal insulator for $m>0$; see Fig.~\ref{fig:FStopology}. In the former phase the band inversion occurs at a momentum $|\vec{k}|_{\rm inv} =\sqrt{|m|/b}$. 

%%%%%%%%%%%%%%%%%%%%%%%%%%%%%%%%%%%%%%%%%%%%%%%%%%%%%%%%%%%%%%%%%%%%%%
%%%%%%%%%%%%%%%%%%%%%%%%%%%%%%%%%%%%%%%%%%%%%%%%%%%%%%%%%%%%%%%%%%%%%%
%%%%%%%%%%%%%%%%%%%%%%%%%%%%%%%%%%%%%%%%%%%%%%%%%%%%%%%%%%%%%%%%%%%%%%
%%%%%%%%%%%%%%%%%%%%%%%%%%%%%%%%%%%%%%%%%%%%%%%%%%%%%%%%%%%%%%%%%%%%%%
%%%%%%%%%%%%%%%%%%%%%%%%%%%%%%%%%%%%%%%%%%%%%%%%%%%%%%%%%%%%%%%%%%%%%%
%%%%%%%%%%%%%%%%%%%%%%%%%%%%%%%%%%%%%%%%%%%%%%%%%%%%%%%%%%%%%%%%%%%%%%
%%%%%%%%%%%%%%%%%%%%%%%%%%%%%%%%%%%%%%%%%%%%%%%%%%%%%%%%%%%%%%%%%%%%%%
%%%%%%%%%%%%%%%%%%%%%%%%%%%%%%%%%%%%%%%%%%%%%%%%%%%%%%%%%%%%%%%%%%%%%%
%%%%%%%%%%%%%%%%%%%%%%%%%%%%%%%%%%%%%%%%%%%%%%%%%%%%%%%%%%%%%%%%%%%%%%
%%%%%%%%%%%%%%%%%%%%%%%%%%%%%%%%%%%%%%%%%%%%%%%%%%%%%%%%%%%%%%%%%%%%%%

The quantum critical point separating two topologically distinct insulating phases is located at $m=0$, where the coefficient of the momentum-dependent Wilson-Dirac mass $b$ flows to zero due to its negative scaling dimension $[b]=-1$ following the general scaling argument. On the other hand, the positive scaling dimension of $m$, namely $[m]=1$, sets the correlation length exponent $\nu=1$ for this quantum phase transition. Although irrelevant, the renormalization group (RG) flow of $b$ stops at the scale of the uniform mass ($m$) inside any insulating phase, and it determines the topology of the gapped states. In this respect, $b$ represents dangerously irrelevant coupling.

Furthermore, when a Dirac insulator is chemically doped ($\mu>0$), a combination of $\mu$, $b$, and $m$ determines the topology of the underlying Fermi surface. Namely, within the topological regime ($m b <0$), when $2 m b<-1$ the system supports an annular Fermi surface for small doping ($\sqrt{-1-4 m b}/(2b)<\mu<m$), whereas a simply connected Fermi surface emerges for larger doping ($\mu>m$) therein. On the other hand, when $2 m b>-1$, which encompasses both topological and normal insulating phases, the Fermi surface is always simply connected~\cite{szaboroy:3DTI}. These features are portrayed in Fig.~\ref{fig:FStopology}.

The existence of an underlying Fermi surface (annular or simply connected) can be conducive to superconductivity even when the microscopic or bare electron-electron interactions are repulsive in nature, following the general principle of the Kohn-Luttinger mechanism~\cite{KL:1, KL:2, KL:3}. Here, we do not delve into the microscopic origin of such repulsive electronic interactions. Nonetheless, in strongly correlated materials, electronic repulsion can arise from Hubbardlike screened Coulomb interaction, whereas in weakly interacting semiconductors effective electron-electron interactions can be sourced by optical phonons below the scale of the optical frequency~\cite{phonon:1, phonon:2, phonon:3}.

In this work, we focus on two representative Dirac insulators in two spatial dimensions, belonging to class A and class AII. From an unbiased leading-order (one-loop) RG calculation we then show that these systems, when chemically doped ($\mu>0$) to sustain a Fermi surface in the normal state, can trigger the nucleation of superconductivity at low temperature, stemming from purely repulsive electronic interactions. In this context, below we offer a synopsis of the central outcomes.

\subsection{Summary of main results}

Here we consider minimal two-band and four-band models for Dirac insulators that belong to class A and class AII, respectively. On analyzing the symmetries of such systems, we arrive at the most general form of local or momentum-independent four-fermion interactions. With the assistance from the Fierz identity, any microscopic model in these systems is shown to be captured by only one and four linearly independent quartic coupling constants, respectively. Performing a leading-order (one loop) RG analysis on such interacting models (see Figs.~\ref{Fig:FeynDiag_Interaction} and~\ref{Fig:FeynDiag_Susceptibility} for the associated Feynman diagrams), controlled by a ``small" parameter $\epsilon=d-1$, we identify the leading instabilities of doped Dirac insulators from the divergence of at least one of such coupling constants at the scale of the underlying Fermi surface. The nature of the ordered state is then unambiguously identified from the simultaneous RG flows of all the symmetry allowed fermion bilinears. Namely, the pattern of symmetry breaking is determined by the finite vacuum expectation value of the fermion bilinear that receives the largest positive correction (anomalous dimension) under coarse graining. Within the framework of this prescription, we construct various cuts of the global phase diagram in the $(g,t)$ plane, where $g$ ($t$) is the four-fermion coupling constant (temperature) that typically display emergent superconductivity at low temperatures. See Figs.~\ref{Fig:QAHIPD}-\ref{fig:QSHIPD3}. In our convention, any bare or microscopic repulsive interaction always corresponds to $g>0$. It should be noted that the transition temperature ($t_c$) predicted from the one-loop RG calculation, performed from the disordered Fermi liquid side, only corresponds to a temperature scale associated with the formation of Cooper pairs. However, a coherent condensation of a macroscopic number of Cooper pairs, leading to superconductivity in the system, occurs at a lower temperature through the Kosterlitz-Thouless transition due to its reduced dimensionality~\cite{KT:original}. Our RG analysis \emph{cannot} capture this phenomenon, which, for example can be achieved from functional RG (fRG) calculations, as has recently been shown for the Hubbard model on a square lattice~\cite{frg:Metzner}.

\begin{figure}[t!]
\includegraphics[width=1.00\linewidth]{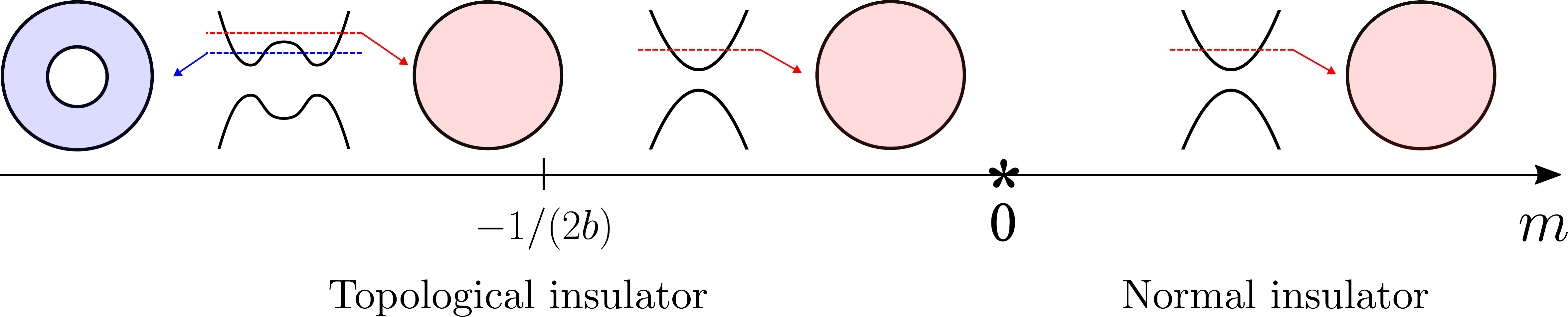}
\caption{Topology of the Fermi surface for various chemical doping, denoted by the dashed colored lines, in a two-dimensional Dirac insulator in the topological ($m<0$) and trivial ($m>0$) regimes with $b>0$. A quantum phase transition between them takes place at $m=0$, around which the Fermi surface is always simply connected, shown by the red circles. For small doping, the Fermi surface becomes annular deep inside the topological regime when $m<-1/(2b)$, shown by the concentric blue circles. Even in this parameter regime, the Fermi surface becomes simply connected for large doping (also shown by a red circle). For $2 m b>-1$ the Fermi surface is always simply connected. Black curves show the energy eigenvalues in the vertical direction along a specific momentum that runs along the horizontal direction.
}~\label{fig:FStopology}
\end{figure}

We note that the nature of the superconducting order, depending on the dominant repulsive interaction in the system, follows certain \emph{selection rules} (discussed in depth in Sec.~\ref{subsec:SROP}). Namely, for a given interaction the corresponding matrix operator, acting on the orbital (in class A and class AII) and spin (only in class AII) degrees of freedom, maximally anticommute with the matrices appearing in the definition of the pairing order parameter. Even though, depending on the parameter values Dirac insulators can be topological or normal and when doped they can support annular or simply connected Fermi surface (see Fig.~\ref{fig:FStopology}), the nature of the superconducting ground state is insensitive to these details as the governing selection rules solely depend on the internal symmetry-based relations (summarized in terms of (anti)commutation relations) among the symmetry class of the system (determining the exact form of the free fermion Hamiltonian), interaction channels, and the paired states. These rules are also operative for the particle-hole or excitonic orders.

Furthermore, to anchor our findings to a concrete microscopic model we construct the phase diagram of repulsive Hubbard models in doped Dirac insulators. In brief, we find that the on-site repulsion operative among the fermions with opposite-parity eigenvalues can be conducive to the nucleation of odd-parity topological $p$-wave pairings in these systems. More specifically, in a class A (class AII) doped Dirac insulator, the on-site orbital or parity Hubbard repulsion favors the condensation of a topological $p+ip$ ($p \pm ip$) pairing that belongs to class D (class DIII) as shown in Fig.~\ref{Fig:QAHIPD} (Fig.~\ref{fig:QSHIPDHubbard}). In turn, such one-to-one correspondences between the symmetry class of the parent or normal state and low-temperature superconducting ground state allows us to draw a comparison with the phenomenological phase diagram in doped magnetic materials, manifesting a confluence of magnetic and pairing orders, which we stage in Fig.~\ref{fig:philosophy}. Emergent topological superconductors, resulting from intra-unit-cell momentum-independent \emph{local} pairings, are energetically superior over extended or momentum-dependent paired states.

\subsection{Organization}

The rest of the paper is organized as follows. In the next section (Sec.~\ref{sec:dopedQAHI}), we discuss the emergent superconductivity in a doped class A Dirac insulator that in the topological parameter regime fosters a quantum anomalous Hall insulator. Section~\ref{sec:dopedQSHI} is devoted to a similar discussion, but for a doped class AII Dirac insulator, which on the other hand yields a quantum spin Hall insulator in the topological regime. Emergent superconductivity within the framework of repulsive Hubbard models in these two systems is discussed in Sec.~\ref{sec:Hubbard}. Our results are summarized in Sec.~\ref{sec:summary}, where we also present discussion on possible material platforms where our theoretical predictions can be experimentally tested. The Fierz reductions of linearly independent four-fermion terms are shown in Appendix~\ref{append:fierz}. Emergent topology resulting from the local superconducting orders in the vicinity of the Fermi surface is shown in Appendix~\ref{append:bandprojection}. The contributions from all the one-loop Feynman diagrams are schematically displayed in Appendix~\ref{append:Feynman}.

%%%%%%%%%%%%%%%%%%%%%%%%%%%%%%%%%%%%%%%%%%%%%%%%%%%%%%%%%%%%%%%%%%%%%%
%%%%%%%%%%%%%%%%%%%%%%%%%%%%%%%%%%%%%%%%%%%%%%%%%%%%%%%%%%%%%%%%%%%%%%
%%%%%%%%%%%%%%%%%%%%%%%%%%%%%%%%%%%%%%%%%%%%%%%%%%%%%%%%%%%%%%%%%%%%%%
%%%%%%%%%%%%%%%%%%%%%%%%%%%%%%%%%%%%%%%%%%%%%%%%%%%%%%%%%%%%%%%%%%%%%%
%%%%%%%%%%%%%%%%%%%%%%%%%%%%%%%%%%%%%%%%%%%%%%%%%%%%%%%%%%%%%%%%%%%%%%
%%%%%%%%%%%%%%%%%%%%%%%%%%%%%%%%%%%%%%%%%%%%%%%%%%%%%%%%%%%%%%%%%%%%%%
%%%%%%%%%%%%%%%%%%%%%%%%%%%%%%%%%%%%%%%%%%%%%%%%%%%%%%%%%%%%%%%%%%%%%%
%%%%%%%%%%%%%%%%%%%%%%%%%%%%%%%%%%%%%%%%%%%%%%%%%%%%%%%%%%%%%%%%%%%%%%
%%%%%%%%%%%%%%%%%%%%%%%%%%%%%%%%%%%%%%%%%%%%%%%%%%%%%%%%%%%%%%%%%%%%%%
%%%%%%%%%%%%%%%%%%%%%%%%%%%%%%%%%%%%%%%%%%%%%%%%%%%%%%%%%%%%%%%%%%%%%%

\section{Doped Dirac Insulator: Class A}~\label{sec:dopedQAHI}

We begin the discussion by promoting the emergence of superconductivity in a doped two-dimensional Dirac insulator that breaks the time-reversal symmetry. The corresponding Hamiltonian assumes the form shown in Eq.~\eqref{eq:DiracUniversal} with $d=2$ and $\Gamma_1=\tau_1$, $\Gamma_2=\tau_2$, and $\Gamma_3=\tau_3$~\cite{QWZ}. The set of Pauli matrices $\{ \tau_\nu\}$ with $\nu=0,1,2,3$ operates on the orbital or parity indices and $\tau_0$ is the two-dimensional identity matrix. The two-component Dirac spinor is given by $\psi^\top_{\vec{k}}=(c_{+},c_{-})(\vec{k})$, where $c_{\tau}(\vec{k})$ is the fermionic annihilation operator on an orbital with parity eigenvalue $\tau=\pm$ and momentum $\vec{k}$. This system describes a collection of spinless or spin-polarized fermions. Depending on the parameter values ($m$ and $b$) the Hamiltonian describes either a quantum anomalous Hall insulator or a normal insulator, and belongs to class A in the ten-fold classification scheme.

%%%%%%%%%%%%%%%%%%%%%%%%%%%%%%%%%%%%%%%%%%%%%%%%%%%%%%%
%%%%%%%%%%%%%%%%%%%%%%%%%%%%%%%%%%%%%%%%%%%%%%%%%%%%%%%
%%%%%%%%%%%%%%%%%%%%%%%%%%%%%%%%%%%%%%%%%%%%%%%%%%%%%%%
%%%%%%%%%%%%%%%%%%%%%%%%%%%%%%%%%%%%%%%%%%%%%%%%%%%%%%%
%%%%%%%%%%%%%%%%%%%%%%%%%%%%%%%%%%%%%%%%%%%%%%%%%%%%%%%
\begin{figure}[t!]
\includegraphics[width=1.00\linewidth]{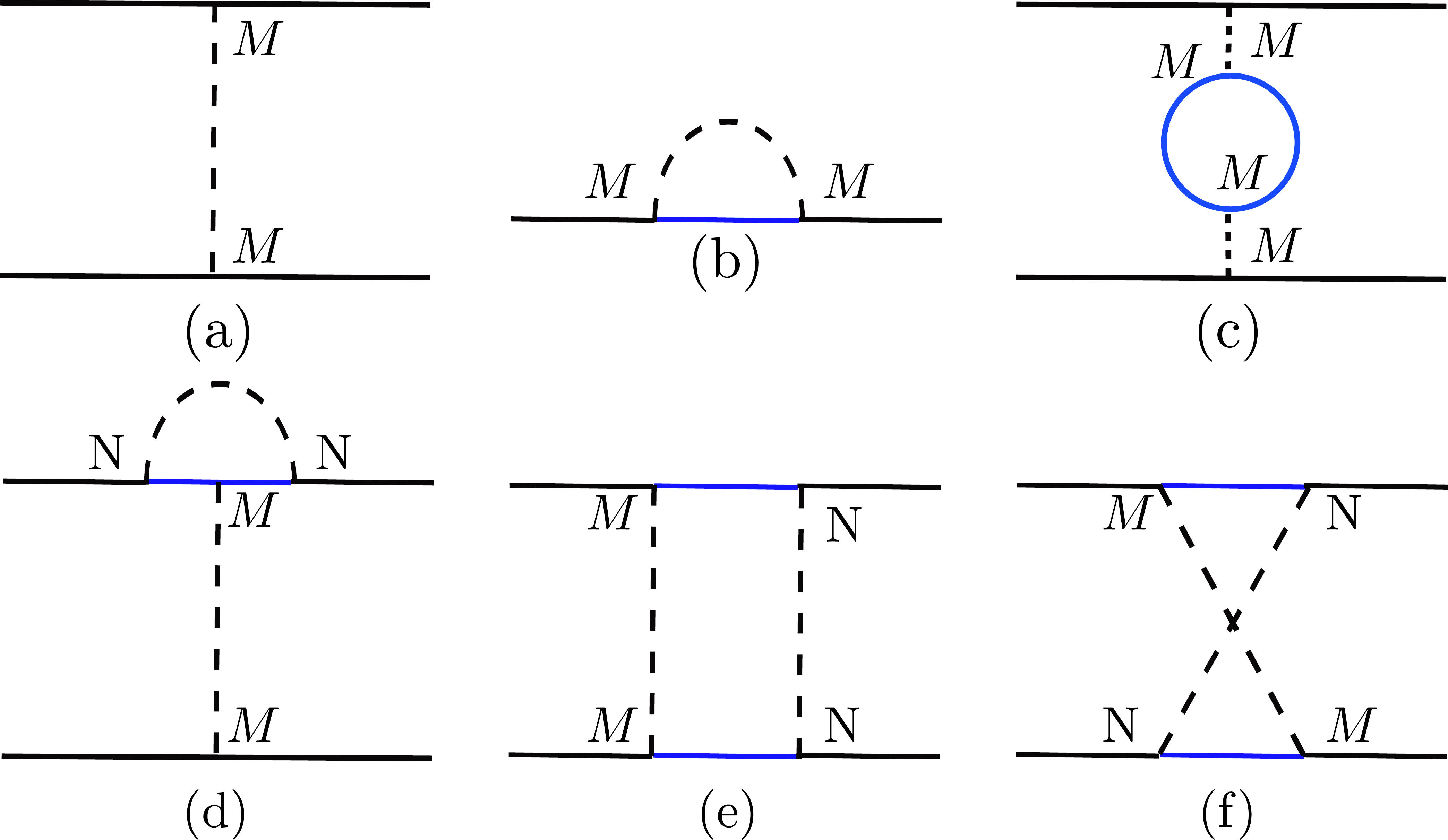}
\caption{(a) The Feynman diagram representing the bare four-fermion interaction $\left( \psi^\dagger M \psi \right)^2$, where $M$ is a Hermitian matrix. (b) Fermionic self-energy correction due to a four-fermions interaction, which is finite only when the chemical potential ($\mu$) is finite, and only renormalizes $\mu$. However, within the framework of a leading-order renormalization group analysis, we ignore such corrections. Corrections to the bare interaction vertex arise from the Feynman diagrams (c)-(f). Here, the solid lines represent fermions, the dashed lines stem from the interaction vertex, and $N$ is a Hermitian matrix as well. The blue lines correspond to the fast modes, living within the thin Wilsonian momentum shell $\Lambda e^{-\ell} <|\vec{k}|<\Lambda$, where $\Lambda$ is the ultraviolet momentum cutoff and $\ell$ is the logarithm of the renormalization group scale. The black lines are the slow modes with momentum $|\vec{k}|<\Lambda e^{-\ell}$.  
}~\label{Fig:FeynDiag_Interaction}
\end{figure}
%%%%%%%%%%%%%%%%%%%%%%%%%%%%%%%%%%%%%%%%%%%%%%%%%%%%%%%
%%%%%%%%%%%%%%%%%%%%%%%%%%%%%%%%%%%%%%%%%%%%%%%%%%%%%%%
%%%%%%%%%%%%%%%%%%%%%%%%%%%%%%%%%%%%%%%%%%%%%%%%%%%%%%%
%%%%%%%%%%%%%%%%%%%%%%%%%%%%%%%%%%%%%%%%%%%%%%%%%%%%%%%
%%%%%%%%%%%%%%%%%%%%%%%%%%%%%%%%%%%%%%%%%%%%%%%%%%%%%%%

This model preserves the parity ($\mathcal P$) symmetry under which $\mathcal{P} \psi_{\vec{k}} \mathcal{P}=\tau_{3} \psi_{-\vec{k}}$, and the charge conjugation symmetry (${\mathcal C}$) under which $\mathcal{C} \psi_{\vec{k}} \mathcal{C}= \tau_1 \psi^\ast_{\vec{k}}$. For concreteness, we also assume that the system possesses a four-fold rotational symmetry ($C^z_4$) about the $z$ axis. Such a rotation is generated by $R^z_{\pi/2}=\exp(i \tau_3 \pi/4)$ under which $(k_x, k_y) \to (k_y,-k_x)$. These symmetries when imposed on the local four-fermion terms, otherwise taking a generic form $g_{_{\nu\rho}} (\psi^\dagger \tau_{\nu} \psi) (\psi^\dagger \tau_{\rho} \psi)$, where $g_{_{\nu\rho}}$ is the corresponding coupling constant, $\nu,\rho=0,1,2,3$, and a summation over repeated indices ($\nu$ and $\rho$) is assumed, severely restrict their form.

The interacting Lagrangian containing all the symmetry allowed local four-fermion terms then takes the form
\begin{equation}~\label{eq:intQAHI}
        L^{\rm A}_{\rm int}= g_{_0} (\psi^\dagger \tau_{0}\psi)^2
				           + g_{_\perp} \sum_{j=1,2} (\psi^\dagger \tau_{j}\psi)^2
									 + g_{_3} (\psi^\dagger \tau_{3} \psi)^2,
\end{equation}
where $\psi \equiv \psi_{\vec{k}}$. The subscripts in the coupling constants denote the projections of the orbital or pseudospin. Namely, the subscript ``0" indicates that the corresponding interaction is operative among pseudospin unpolarized fermions, subscript ``$\perp$" (``3") indicates that the corresponding quartic interaction takes place among fermions with pseudospin aligned in the easy-plane (easy-axis) or the $xy$ plane ($z$ axis). However, not all three quartic terms are linearly independent due to the Fierz relations among them~\cite{Fierz:1, Fierz:2, Fierz:3, Fierz:4, Fierz:5}. As shown in Appendix~\ref{append:fierz}, only one quartic term is linearly independent, which we choose to be the one proportional to $g_{_0}$ without any loss of generality. The remaining two quartic term can be expressed in terms of $(\psi^\dagger \tau_{0}\psi)^2$. Then the imaginary time ($\tau$) Euclidean action can be decomposed as $S=S_0 + S_{\rm int}$, where 
\allowdisplaybreaks[4]
\begin{eqnarray}~\label{eq:actionInt}
S_0 &=& \int d\tau \int d^d\vec{x} \; \psi^\dag_{\tau,\vec{x}} \left[ \partial_\tau + \hat{h}_{\rm Dir}(\vec{k} \to -i\boldsymbol{\nabla}) \right]\psi_{\tau,\vec{x}} \;, \nonumber \\
S_{\rm int} &=& \int d\tau \int d^d\vec{x} \; g_{_0} \left( \psi^\dag_{\tau,\vec{x}} \tau_0 \psi_{\tau,\vec{x}} \right)^2, 
\end{eqnarray}   
and $\Psi^\dag_{\tau,\vec{x}}$ and $\Psi_{\tau,\vec{x}}$ are independent Grassmann variables. We emphasize that any microscopic model of local interactions can be captured only in terms of $g_{_0}$, which we exemplify with a Hubbard model in this system in Sec.~\ref{sec:Hubbard}.

The scale invariance of $S_0$ implies the scaling dimension of the fermionic fields is $[\psi]=[\psi^\dagger]=d/2$, that of momentum is $[\vec{k}]=1$ and the fermionic Matsubara frequency is $[\omega_n]=z=1$. The fermionic Matsubara frequency is given by $\omega_n= (2n+1)\pi T$, where $T$ denotes the temperature and $n \in (-\infty, \infty)$ is an integer, yielding $[T]=z$. The scaling dimension of the uniform mass is $[m]=z$ which is a relevant parameter. The scaling dimension of the coefficient of the momentum-dependent Wilson-Dirac mass $[b]=z-2=-1$ in a Dirac system with $z=1$ and thus $b$ is an irrelevant parameter. The scaling dimension of the chemical potential is $[\mu]=z$. Then the scaling dimension of the four-fermion coupling constant $g_{_0}$ is given by $[g_{_0}]=z-d$ (set by the scale invariance of $S_{\rm int}$), which is thus irrelevant in a Dirac system with $z=1$ in two dimensions ($d=2$). Hence, any ordering can only set in beyond a threshold strength of the four-fermion interaction that can be shown from a controlled $\epsilon$ expansion about the lower-critical one spatial dimension, where they becomes \emph{marginal}~\cite{grossneveu}, with $\epsilon=d-1$~\cite{epsilon:1, epsilon:2}.

%%%%%%%%%%%%%%%%%%%%%%%%%%%%%%%%%%%%%%%%%%%%%%%%%%%%%%%
%%%%%%%%%%%%%%%%%%%%%%%%%%%%%%%%%%%%%%%%%%%%%%%%%%%%%%%
%%%%%%%%%%%%%%%%%%%%%%%%%%%%%%%%%%%%%%%%%%%%%%%%%%%%%%%
%%%%%%%%%%%%%%%%%%%%%%%%%%%%%%%%%%%%%%%%%%%%%%%%%%%%%%%
%%%%%%%%%%%%%%%%%%%%%%%%%%%%%%%%%%%%%%%%%%%%%%%%%%%%%%%
\begin{table}[t!]
\begin{tabular}{|c|c|c|c|c|c|}
\hline
CF & Bilinears & Physical meaning & ${\mathcal P}$ & ${\mathcal C}$ & $C^z_4$ \\
\hline \hline
$\Delta_0$     & $\psi^{\dagger} \eta_3\tau_{0}\psi$      & Charge density                & $+$ & $-$ & $0$ \\
$\Delta_\perp$ & $\psi^{\dagger} \eta_3\tau_{j}\psi$      & Abelian current               & $-$ & $-$ & $1$ \\
$\Delta_3$     & $\psi^{\dagger} \eta_0\tau_{3}\psi$      & Symmetric Dirac mass          & $+$ & $+$ & $0$ \\
$\Delta_p$     & $\psi^{\dagger} \eta_\alpha\tau_{3}\psi$ & Isotropic odd-parity pairing  & $-$ & $+$ & $0$ \\
\hline
\end{tabular}
\caption{Momentum-independent local excitonic (first three rows) and superconducting (last row) orders in a class A system with their conjugate fields (CFs) are shown in the first column, see Sec.~\ref{sec:dopedQAHI}. The corresponding fermion bilinears in the Nambu-doubled basis are shown in the second column, where $\psi \equiv \psi_{\rm Nam}$ (for brevity). The physical meaning of each fermion bilinear is mentioned in the third column. Transformations of each fermion bilinear under the parity (${\mathcal P}$), charge conjugation (${\mathcal C}$), and fourfold rotation about the $z$ axis ($C^z_4$) are shown in the fourth, fifth, and sixth columns, respectively. Here $+$ ($-$) indicates even (odd) and $0$ ($1$) corresponds to scalar (vector). For the pairing order $\alpha=1,2$, reflecting the gauge redundancy in defining its U(1) phase.
}~\label{tab:ClassA}
\end{table}
%%%%%%%%%%%%%%%%%%%%%%%%%%%%%%%%%%%%%%%%%%%%%%%%%%%%%%%
%%%%%%%%%%%%%%%%%%%%%%%%%%%%%%%%%%%%%%%%%%%%%%%%%%%%%%%
%%%%%%%%%%%%%%%%%%%%%%%%%%%%%%%%%%%%%%%%%%%%%%%%%%%%%%%
%%%%%%%%%%%%%%%%%%%%%%%%%%%%%%%%%%%%%%%%%%%%%%%%%%%%%%%
%%%%%%%%%%%%%%%%%%%%%%%%%%%%%%%%%%%%%%%%%%%%%%%%%%%%%%%

As we are interested in capturing the nature of the ordered states, although mainly superconductivity, besides the RG flow equation for $g_{_0}$, we also take into account the RG flow equations for the source terms associated with various fermion bilinears. The imaginary time action containing all the local or momentum-independent excitonic (exc) or particle-hole and pairing (pair) or particle-particle orders takes the form 
\begin{equation}~\label{eq:actionSource}
S_{\rm source}= \int d\tau \int d^d\vec{x} \;  \left( h^{\rm A}_{\rm exc} + h^{\rm A}_{\rm pair} \right). 
\end{equation}   
In order to capture all the orders in the same framework, we now introduce a Nambu-doubled spinor basis $\psi^\top_{\rm Nam}=(\psi_{\omega_n,k}, \tau_{1}\psi_{-\omega_n,-k}^\star)$ with $\star$ denoting the complex conjugate. In this basis
\begin{equation}~\label{eq:hamexcQAHI}
    h^{\rm A}_{\rm exc} =\Delta_0 (\Psi^\dagger \Gamma_{30}\psi)
		        + \Delta_\perp \sum_{j=1,2} (\Psi^\dagger\Gamma_{3j}\psi) 
						+ \Delta_3 (\Psi^\dagger\Gamma_{03}\psi),
\end{equation}
where $\psi \equiv \psi_{\rm Nam}$ (used for brevity) and the four-component Hermitian matrix $\Gamma_{\nu \rho}=\eta_\nu \otimes \tau_\rho$. The newly introduced set of Pauli matrices $\{ \eta_\nu \}$ operate on the Nambu or particle-hole index and $\otimes$ denote the tensor product. This system permits only one local pairing (the number of two-dimensional imaginary Hermitian matrices due to the Pauli exclusion principle~\cite{roynevidomskyy}), leading to 
\begin{equation}~\label{eq:hampairQAHI}
        h^{\rm A}_{\rm pair}=\Delta_p \; \sum_{\alpha=1,2} (\psi^\dagger \Gamma_{\alpha 3}\psi),
\end{equation}
where $\alpha=1,2$ manifests the U($1$) gauge redundancy in defining the pairing order parameter. Close to the Fermi surface this local pairing assumes the form of a topological $p+ip$ pairing as shown in Appendix~\ref{append:bandprojection}. Transformations of all the fermion bilinears under various symmetries of the noninteracting system are summarized in Table~\ref{tab:ClassA}, where we also mention their
physical nature. Due to the broken time-reversal symmetry in the normal state, a class A system does not support any $p \pm ip$ pairing and only fosters $p+ip$ pairing. If we sacrifice the fourfold rotational symmetry in the normal state then the resulting $p+ip$ pairing also breaks such a symmetry, yielding a nematic $p+ip$ pairing without altering its topological properties as this paired state trasforms as a scalar under fourfold rotations (see Table~\ref{tab:ClassA}).

In the announced Nambu-doubled basis the Dirac Hamiltonian takes the explicit form 
\begin{equation}~\label{HDirNamA}
\hat{h}^{\rm Nam}_{\rm Dir, A}= v \left( \Gamma_{01} k_x + \Gamma_{02} k_y \right) + (m+bk^2) \Gamma_{03}-\mu\Gamma_{30}.
\end{equation}
Accordingly, the fermionic Green's function in this basis is given by 
\begin{equation}~\label{eq:GreensNambu}
G_{\rm fer} = G_+ \oplus G_-,
\end{equation}
where with $\Omega_\pm = i\omega_n \pm \mu$ and $E_k=\sqrt{v^2 k^2+ (m+b k^2)^2}$,
\begin{equation}
G_\pm = \frac{\Omega_\pm \tau_0 + v (\tau_1 k_x + \tau_2 k_y) + (m+b k^2) \tau_3}{\Omega^2_\pm-E^2_k}.
\end{equation}
Finally note that in the Nambu basis, the matrix operator appearing in the four-fermion term $\tau_0 \to \Gamma_{30}$.

%%%%%%%%%%%%%%%%%%%%%%%%%%%%%%%%%%%%%%%%%%%%%%%%%%%%%%%
%%%%%%%%%%%%%%%%%%%%%%%%%%%%%%%%%%%%%%%%%%%%%%%%%%%%%%%
%%%%%%%%%%%%%%%%%%%%%%%%%%%%%%%%%%%%%%%%%%%%%%%%%%%%%%%
%%%%%%%%%%%%%%%%%%%%%%%%%%%%%%%%%%%%%%%%%%%%%%%%%%%%%%%
%%%%%%%%%%%%%%%%%%%%%%%%%%%%%%%%%%%%%%%%%%%%%%%%%%%%%%%
\begin{figure}[t!]
\includegraphics[width=1.00\linewidth]{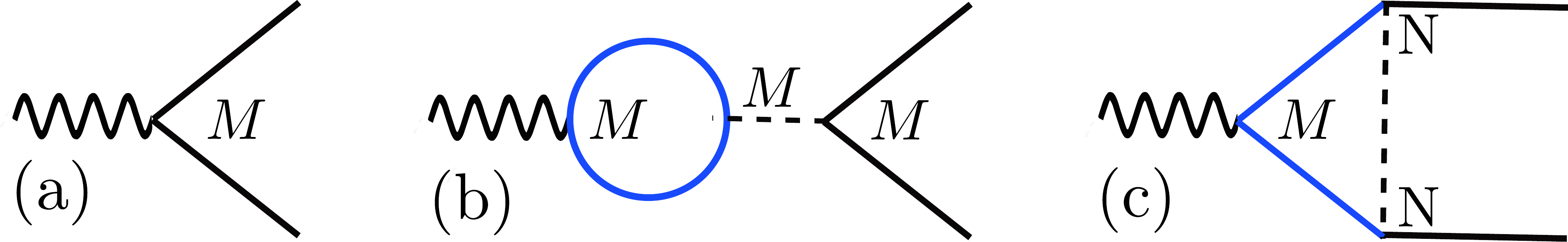}
\caption{(a) Feynman diagram representing the bare vertex associated with the source term $\psi^\dagger M \psi$. The leading order renormalization of such vertices arises from the Feynman diagrams in (b) and (c). Here, wavy lines stand for the source field, solid lines for fermions, and the dashed lines stem from the interaction vertex. The blue (black) solid lines represent fast (slow) modes. Here $M$ and $N$ are Hermitian matrices.  
}~\label{Fig:FeynDiag_Susceptibility}
\end{figure}
%%%%%%%%%%%%%%%%%%%%%%%%%%%%%%%%%%%%%%%%%%%%%%%%%%%%%%%
%%%%%%%%%%%%%%%%%%%%%%%%%%%%%%%%%%%%%%%%%%%%%%%%%%%%%%%
%%%%%%%%%%%%%%%%%%%%%%%%%%%%%%%%%%%%%%%%%%%%%%%%%%%%%%%
%%%%%%%%%%%%%%%%%%%%%%%%%%%%%%%%%%%%%%%%%%%%%%%%%%%%%%%
%%%%%%%%%%%%%%%%%%%%%%%%%%%%%%%%%%%%%%%%%%%%%%%%%%%%%%%

The Feynman diagrams responsible for the leading-order renormalization of the quartic term $g_{_0}$ are shown in Fig.~\ref{Fig:FeynDiag_Interaction}. The resulting RG flow equation in terms of the dimensionless coupling constant defined as $2 \pi g_{_0} \Lambda^\epsilon/v \to g_{_0}$ reads as (see Appendix~\ref{append:Feynman} for details)
\begin{equation}~\label{eq:RGcouplingQAHI}   
	\beta_{g_{_0}}= \frac{dg_{_0}}{d \ell}=- \epsilon g_{_{0}}-2(f_k-\tilde f_k+f_0+\tilde f_0-f_w+\tilde f_w) g_{_{0}}^2,
\end{equation}
obtained after performing the summation over the fermionic Matsubara frequencies and integrating out a thin Wilsonian momentum shell with $\Lambda e^{-\ell} < |\vec{k}|<\Lambda$, and subsequently recasting $S_{\rm int}$ in its original form but in terms of a scale-dependent coupling constant $g_{_0}(\ell)$~\cite{Wilson:1, Wilson:2, Wilson:3}. Here, $\Lambda \sim 1/a$ is the ultraviolet momentum cutoff up to which the energy dispersion scales linearly with momentum, $a$ bears the dimension of the lattice constant, and $\ell$ is the logarithm of the RG scale.

%%%%%%%%%%%%%%%%%%%%%%%%%%%%%%%%%%%%%%%%%%%%%%%%%%%%%%%%%%%%%%%%%%%%%%
%%%%%%%%%%%%%%%%%%%%%%%%%%%%%%%%%%%%%%%%%%%%%%%%%%%%%%%%%%%%%%%%%%%%%%
%%%%%%%%%%%%%%%%%%%%%%%%%%%%%%%%%%%%%%%%%%%%%%%%%%%%%%%%%%%%%%%%%%%%%%
%%%%%%%%%%%%%%%%%%%%%%%%%%%%%%%%%%%%%%%%%%%%%%%%%%%%%%%%%%%%%%%%%%%%%%
%%%%%%%%%%%%%%%%%%%%%%%%%%%%%%%%%%%%%%%%%%%%%%%%%%%%%%%%%%%%%%%%%%%%%%
\begin{figure*}[t!]
\includegraphics[width=1.00\linewidth]{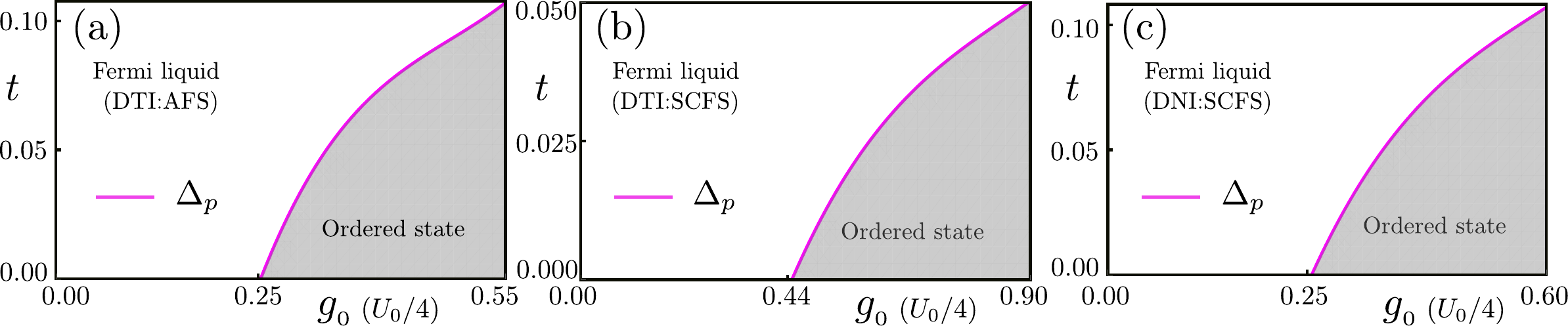}
\caption{Cuts of the global phase diagram for repulsive density-density interactions ($g_{_0}$) in a time-reversal symmetry breaking doped planar Dirac insulator from class A for (a) $m=-0.9$, $b=1.0$, and $\mu=0.86$, (b) $m=-0.9$, $b=1.0$, and $\mu=0.91$, and (c) $m=0.05$, $b=0.05$, and $\mu=0.86$. The horizontal (vertical) axis corresponds to dimensionless coupling constant (temperature). The white (shaded) regions correspond to disordered Fermi liquid (ordered state). The nature of the ordered state is color coded along the phase boundary between the ordered and disordered phases, which also yields the dimensionless transition temperature ($t_c$). In this case the ordered phase always represents the only local paired state allowed in this system ($\Delta_p$), which around the Fermi surface assumes the form of a topological $p+ip$ pairing (see Appendix~\ref{append:bandprojection}). The abbreviation DTI (DNI) stands for doped topological (normal) insulator, indicating the nature of the normal state in the absence of any doping ($\mu=0$), while AFS (SCFS) stands for annular (simply connected) Fermi surface when $\mu$ is finite. See Sec.~\ref{sec:dopedQAHI} for details. The phase diagrams with repulsive on-site orbital Hubbard repulsion ($U_0$) are identical in which the interaction axis is $U_0/4$, see Sec.~\ref{sec:Hubbard}. Dimensionless coupling constants ($g_{_0}$ and $U_0$) are measured in units of $\epsilon$, where $\epsilon=d-1$.  
}~\label{Fig:QAHIPD}
\end{figure*}
%%%%%%%%%%%%%%%%%%%%%%%%%%%%%%%%%%%%%%%%%%%%%%%%%%%%%%%%%%%%%%%%%%%%%%
%%%%%%%%%%%%%%%%%%%%%%%%%%%%%%%%%%%%%%%%%%%%%%%%%%%%%%%%%%%%%%%%%%%%%%
%%%%%%%%%%%%%%%%%%%%%%%%%%%%%%%%%%%%%%%%%%%%%%%%%%%%%%%%%%%%%%%%%%%%%%
%%%%%%%%%%%%%%%%%%%%%%%%%%%%%%%%%%%%%%%%%%%%%%%%%%%%%%%%%%%%%%%%%%%%%%
%%%%%%%%%%%%%%%%%%%%%%%%%%%%%%%%%%%%%%%%%%%%%%%%%%%%%%%%%%%%%%%%%%%%%%

Besides the flow of $g_{_0}$ we also need to consider the RG flows of the following dimensionless parameters, defined as $T/(\Lambda v) \to t$, $\mu/(\Lambda v) \to \mu$, $m/(\Lambda v) \to m $, and $b \Lambda/v \to b$. They are given by 
\begin{equation}~\label{eq:Parameters}
\frac{dt}{d \ell}=z t, \; 
\frac{d\mu}{d \ell}= z \mu, \;
\frac{dm}{d \ell}= z m, 
\:\: \text{and} \:\:
\frac{db}{d \ell}= (z-2)b,
\end{equation}   
respectively, with $z=1$ in a Dirac system~\cite{RGtemp:1, RGtempt:2}. Notice that the last set of flow equations follows from the scaling dimension of the corresponding quantity at the bare level.

Feynman diagrams yielding the leading-order RG flow equations for the source terms are shown in Fig.~\ref{Fig:FeynDiag_Susceptibility} and they are given by (see Appendix~\ref{append:Feynman} for details)
\allowdisplaybreaks[4]
\begin{eqnarray}~\label{eq:RGsource}
\bar{\beta}_{\Delta_{0}} &=& \frac{d \ln \Delta_{0}}{d\ell}-z= -2(2f_k+f_0+f_w) g_{_{0}}, \nonumber \\
\bar{\beta}_{\Delta_{\perp}} &=& \frac{d \ln \Delta_{\perp}}{d\ell}-z = -2(f_0-f_w) g_{_{0}}, \nonumber \\
\bar{\beta}_{\Delta_{3}} &=& \frac{d \ln \Delta_{3}}{d\ell}-z= -2(2f_k-f_0-f_w) g_{_{0}}, \nonumber \\
\text{and} \: \bar{\beta}_{\Delta_{p}} &=& \frac{d \ln \Delta_{p}}{d\ell}-z= -2(-2\tilde f_k+\tilde f_0+\tilde f_w) g_{_{0}}. 
\end{eqnarray}
Various functions appearing in the RG flow equations are explicitly given by
\allowdisplaybreaks[4]
\begin{eqnarray}~\label{eq:flowfunctions}
f_k &=& \frac{1}{2\pi} \frac{\Lambda^d}{v \Lambda} \: \sum_{\tau=\pm} \left[ \frac{\tanh\left(\frac{E_{\Lambda}+\tau \mu}{2t}\right)}{8 E_{\Lambda}^3 } - \frac{ \sech^2\left(\frac{E_{\Lambda}+\tau \mu}{2t}\right)}{16 E_{\Lambda}^2 t} \right], \nonumber \\
	%%%%%%%%%%%%%%%%%%%%%%%%%%%%%%%%%%%%%%%%%%%%%%%%%%%%%%%%%%%%%%%%%%%%%%
\tilde f_k &=& \frac{1}{2\pi} \frac{\Lambda^d}{v \Lambda} \: \sum_{\tau=\pm} \tau \frac{\tanh\left(\frac{E_{\Lambda}-\tau \mu}{2t }\right) \left( E_{\Lambda}+\tau \mu \right)}{8 E_{\Lambda} \mu \left( E_{\Lambda}^2-\mu^2 \right)}, \nonumber \\
%%%%%%%%%%%%%%%%%%%%%%%%%%%%%%%%%%%%%%%%%%%%%%%%%%%%%%%%%%%%%%%%%%%%%%
f_0 &=& \frac{1}{2\pi} \frac{\Lambda^d}{v \Lambda}(m+b)^2 \nonumber \\
&\times& \sum_{\tau=\pm}\left[ \frac{\tanh\left(\frac{E_{\Lambda}+\tau \mu}{2t }\right)}{8 E_{\Lambda}^3}
	- \frac{\sech^2\left(\frac{E_{\Lambda}+\tau \mu}{2t}\right)}{16 E_{\Lambda}^2 t} \right], \nonumber \\
	%%%%%%%%%%%%%%%%%%%%%%%%%%%%%%%%%%%%%%%%%%%%%%%%%%%%%%%%%%%%%%%%%%%%%%
{\tilde f}_0 &=& \frac{1}{2\pi} \frac{\Lambda^d}{v \Lambda}(m+b)^2 \sum_{\tau=\pm} \tau \frac{\tanh\left(\frac{E_{\Lambda}-\tau \mu}{2t}\right)\left(E_{\Lambda}+\tau \mu\right)}{8 E_{\Lambda} \mu \left(E_{\Lambda}^2-\mu^2\right)}, \nonumber\\
%%%%%%%%%%%%%%%%%%%%%%%%%%%%%%%%%%%%%%%%%%%%%%%%%%%%%%%%%%%%%%%%%%%%%%
f_w &=& -\frac{1}{2\pi} \frac{\Lambda^d}{v \Lambda} \sum_{\tau=\pm} \left[\frac{\sech^2\left(\frac{E_{\Lambda}+\tau \mu}{2t }\right)}{16t } +\frac{\tanh\left(\frac{E_{\Lambda}+\tau \mu}{2t}\right)}{8 E_{\Lambda}} \right], \nonumber \\
	%%%%%%%%%%%%%%%%%%%%%%%%%%%%%%%%%%%%%%%%%%%%%%%%%%%%%%%%%%%%%%%%%%%%%%
\text{and} \nonumber \\
\tilde f_w &=& \frac{1}{2\pi} \frac{\Lambda^d}{v \Lambda} \nonumber \\
&\times& \sum_{\tau=\pm} \tau \frac{\tanh\left(\frac{E_{\Lambda}-\tau \mu}{2t}\right) \left( E_{\Lambda}^2-\tau  E_{\Lambda} \mu- 2\mu^2 \right)}{8  \mu \left( E_{\Lambda}^2-\mu^2 \right)},
\end{eqnarray}
where $E_{\Lambda}=[1+(m+b)^2]^{1/2}$.

At this stage a comment is due. Although, we consider the RG flow of the conjugate field $\Delta_0$ that couples with the fermion density, the corresponding fermion bilinear does not break any microscopic symmetry, besides the charge conjugation. The renormalized value of $\Delta_0$ should then be included in the RG flow of the chemical potential ($\mu$), shown in Eq.~\eqref{eq:Parameters}. However, $\mu$ enters the RG flow equation of the coupling constant(s) through the functions defined in Eq.~\eqref{eq:flowfunctions}, each of which corresponds to the coefficient of a quadratic function of the coupling constant(s). Hence, to maintain the order of the perturbative expansion of $S_{\rm int}$ we neglect the RG corrections of $\Delta_0$ resulting from the Feynman diagrams in Fig.~\ref{Fig:FeynDiag_Susceptibility} or equivalently any perturbative correction to $\mu$, stemming from the Feynman diagram in Fig.~\ref{Fig:FeynDiag_Interaction}(b), when the RG flow equation(s) of the coupling constant(s) is (are) computed to the quadratic or one-loop order; see Eq.~\eqref{eq:RGcouplingQAHI}. We follow this strategy in Sec.~\ref{sec:dopedQSHI}. Next we numerically solve the coupled sets of RG flow equations to underpin the ordered states and construct various cuts of the global phase diagram of interacting doped class A Dirac insulator.

\subsection{Ordered states and phase diagrams}

For any bare value of $g_{_0}$ at the scale of the ultraviolet cutoff $\Lambda$ or $\ell=0$, denoted by $g_{_0}(0)$, the RG flow of $g_{_0}$  has to be stopped at an infrared scale $\ell^\star_\mu=-\ln(\mu(0))/z$ in the presence of an underlying Fermi surface, where $\mu(0)<1$ is the bare value of the dimensionless chemical potential $\mu$~\cite{szaboroy:3DTI, finitemu:1, finitemu:2, finitemu:3, finitemu:4, finitemu:5}. Notice that the RG scale $\ell^\star_\mu$ is set by the underlying Fermi surface and all the states below it are completely filled, among which no inelastic scattering (due to four-fermion interactions) can take place. Thus all the RG flows must be stopped at this scale. In the disordered phase (without any symmetry breaking) $g_{_0}(\ell)$ decreases under coarse graining (with increasing $\ell$) and it does not diverge. The onset of an ordered phase is characterized by the divergence of $g_{_0}(\ell)$ at an RG scale $\ell_{\rm div} =\ell^\star_\mu$. On the other hand, within the ordered phase this coupling constant diverges at an RG scale $\ell_{\rm div}<\ell^\star_\mu$. The exact nature of the leading instability or the pattern of symmetry breaking is then unambiguously identified by the renormalized susceptibility $\Delta_\beta(\ell)$ that receives the largest positive correction (anomalous dimension) when $g_{_0}(\ell)$ diverges at the RG scale $\ell_{\rm div}$, where $\beta=0,\perp,3,p$. Throughout, we set the bare temperature $t(0) =5 \times 10^{-6}$ to identify the dominant instability in a doped time-reversal symmetry breaking Dirac insulator close to the zero temperature for repulsive interactions in various channel.

We find that only density-density repulsive interaction with $g_{_0}(0)>0$ yields superconductivity, irrespective of the geometry of the underlying Fermi surface (annular or simply connected). So, we solely focus on this interaction channel for the rest of this section. The superconducting ground state in this case is unique, denoted by $\Delta_p$. The associated reduced BCS Hamiltonian obtained by projecting this paired state on the conduction or valence band fostering the Fermi surface assumes the form of a time-reversal symmetry breaking topological $p+ip$ pairing, as shown in Appendix~\ref{append:bandprojection}. Thus, close to zero temperature the repulsive density-density interaction in this system accommodates a topological superconductor. The remaining two interactions, namely $g_{_\perp}$ and $g_{_3}$ [see Eq.~\eqref{eq:intQAHI}] when repulsive, do not yield any superconductivity in the system at any temperature and only support excitonic or particle-hole orders. For this reason, we do not show the phase diagrams in the $(g_{_\perp}, t)$ or $(g_{_3}, t)$ plane.

%%%%%%%%%%%%%%%%%%%%%%%%%%%%%%%%%%%%%%%%%%%%%%%%%%%%%%%%%%%%%%%%%%%%%%%%%%%%%%%%%%%%%%%%%%%%%%%
%%%%%%%%%%%%%%%%%%%%%%%%%%%%%%%%%%%%%%%%%%%%%%%%%%%%%%%%%%%%%%%%%%%%%%%%%%%%%%%%%%%%%%%%%%%%%%%
%%%%%%%%%%%%%%%%%%%%%%%%%%%%%%%%%%%%%%%%%%%%%%%%%%%%%%%%%%%%%%%%%%%%%%%%%%%%%%%%%%%%%%%%%%%%%%%
%%%%%%%%%%%%%%%%%%%%%%%%%%%%%%%%%%%%%%%%%%%%%%%%%%%%%%%%%%%%%%%%%%%%%%%%%%%%%%%%%%%%%%%%%%%%%%%
%%%%%%%%%%%%%%%%%%%%%%%%%%%%%%%%%%%%%%%%%%%%%%%%%%%%%%%%%%%%%%%%%%%%%%%%%%%%%%%%%%%%%%%%%%%%%%%
\begin{table*}[t!]
\begin{tabular}{|c|c|c|c|c|c|c|c|}
\hline
CF & Matrix & Physical meaning & ${\mathcal T}$ & ${\mathcal P}$ & ${\mathcal C}$ & $C^z_4$ & $S^z_4$  \\
\hline \hline
$\Delta_1^{p}$ &  $(\eta_1, \eta_2) \; \Gamma_{00}$ & $s$-wave pairing & $(+,-)$ & $+$ & $+$ & 0 & $0$ \\
$\Delta_2^{p}$ &  $(\eta_1, \eta_2) \; \Gamma_{0j}$ & Nematic odd-parity pairing & $(+,-)$ & $-$ & $+$ &1 & $0$  \\
$\Delta_3^{p}$ &  $(\eta_1, \eta_2) \; \Gamma_{33}$ & $s$-wave pairing (gap $\sim m$)  & $(+,-)$ & $+$ & $+$ &0 & $0$  \\
$\Delta_4^{p}$ &  $(\eta_1, \eta_2) \; \Gamma_{a3}$ & Isotropic odd-parity pairing & $(+,-)$ & $-$ & $+$ &0 & $1$  \\
\hline \hline
$\Delta_0^{s}$ &  $\eta_3\Gamma_{00}$ & Fermionic density & $+$ & $+$ & $-$   & $0$ & $0$ \\
$\Delta_{\perp}^{s}$ &  $\eta_3\Gamma_{0j}$ & Spin current & $+$ & $-$ & $-$  & $1$ & $0$ \\
$\Delta_3^{s}$ &  $\eta_0\Gamma_{03}$ & Staggered orbital density & $-$ & $+$ & $+$  & $0$ & $0$   \\
$\Delta_0^{z}$ &  $\eta_0\Gamma_{30}$ & Easy-axis spin density & $-$ & $+$ & $-$ & $0$ & $0$  \\
$\Delta_{\perp}^{z}$ &  $\eta_0\Gamma_{3j}$ & Abelian Current & $-$ & $-$ & $-$ & $1$ & $0$ \\
$\Delta_3^{z}$ &  $\eta_3\Gamma_{33}$ &  Symmetric Dirac mass & $+$ & $+$ & $+$ & $0$ & $0$ \\
$\Delta_0^{\perp}$ &  $\eta_0 (\Gamma_{10}, \Gamma_{20})$ & Easy-plane spin density & $-$ & $-$ & $(-,+)$  & $0$ & $1$ \\
$\Delta_{\perp}^{\perp}$ &  $\eta_0 (\Gamma_{1j}, \Gamma_{2j})$ & Easy-plane spin current & $-$ & $+$ & $(-,+)$  & $1$ & $1$ \\
$\Delta_{3}^{\perp}$ &  $\eta_3(\Gamma_{13},\Gamma_{23})$ & Easy-plane staggered orbital density & $+$ & $-$ & $(+,-)$ & $0$ & $1$  \\
\hline
\end{tabular}
\caption{Momentum-independent local superconducting (first four rows) and excitonic (last nine rows) orders with their conjugate fields (CFs) in a class AII system are shown in the first column, see Sec.~\ref{sec:dopedQSHI}. The corresponding matrix operators associated with the fermion bilinears $\psi^\dag_{\rm Nam} \eta_\mu \Gamma_{\nu \rho} \psi_{\rm Nam}$ in the Nambu-doubled basis are shown in the second column. The physical meaning of each order is displayed in the third column. Here, $j=1,2$ indicate that the pseudospin degrees of freedom is projected onto the easy-plane and $a=1,2$ mark that the spin degrees of freedom is confined within the easy-plane. Transformations of all the fermion bilinears under the discrete time-reversal (${\mathcal T}$), parity (${\mathcal P}$), and charge conjugation (${\mathcal C}$) symmetries are shown in the fourth, fifth, and sixth columns, respectively. In the seventh and eight columns we display the transformations of all the fermion bilinears under the fourfold rotation about the $z$ axis ($C^z_4$) and the fourfold rotation of the spin quantization axis about the $z$ direction ($S^z_4$), respectively. Here $+$ ($-$) indicates even (odd) and $0$ ($1$) corresponds to scalar (vector). The easy-plane (easy-axis) corresponds to the $xy$ plane ($z$ axis). Notice that the easy-plane (easy-axis) characterization of a fermion bilinear manifests the spin orientation therein in the $xy$ plane ($z$ direction).    
}~\label{tab:bilinears}
\end{table*}
%%%%%%%%%%%%%%%%%%%%%%%%%%%%%%%%%%%%%%%%%%%%%%%%%%%%%%%%%%%%%%%%%%%%%%%%%%%%%%%%%%%%%%%%%%%%%%%
%%%%%%%%%%%%%%%%%%%%%%%%%%%%%%%%%%%%%%%%%%%%%%%%%%%%%%%%%%%%%%%%%%%%%%%%%%%%%%%%%%%%%%%%%%%%%%%
%%%%%%%%%%%%%%%%%%%%%%%%%%%%%%%%%%%%%%%%%%%%%%%%%%%%%%%%%%%%%%%%%%%%%%%%%%%%%%%%%%%%%%%%%%%%%%%
%%%%%%%%%%%%%%%%%%%%%%%%%%%%%%%%%%%%%%%%%%%%%%%%%%%%%%%%%%%%%%%%%%%%%%%%%%%%%%%%%%%%%%%%%%%%%%%
%%%%%%%%%%%%%%%%%%%%%%%%%%%%%%%%%%%%%%%%%%%%%%%%%%%%%%%%%%%%%%%%%%%%%%%%%%%%%%%%%%%%%%%%%%%%%%%

Even though we set the bare dimensionless temperature $t(0) =5 \times 10^{-6}$ to identify the dominant instability of a Fermi liquid state, obtained by doping a time-reversal symmetry breaking planar Dirac insulator, sufficiently close to the zero temperature, the notion of $\ell_{\rm div}$ provides a semiquantitative estimation of the transition temperature ($t_c$) in the following way. The solution of the RG flow equation for dimensionless temperature from Eq.~\eqref{eq:Parameters} reads as $t(\ell)=t(0)\exp(z \ell)$. In an interacting doped Dirac insulator, there are two infrared RG scales $\ell^\star_\mu$ and $\ell_{\rm div}$, respectively at which the renormalized chemical potential and the coupling constant $g_{_0}$ becomes of the order of \emph{unity}. Associated with these two RG scales, we define two bare temperatures $t(0)=t_\mu$ and $t_g$, respectively, such that $t(\ell^\star_\mu)=t(\ell_{\rm div})=1$, yielding $t_\mu=\exp[-z \ell^\star_\mu]$ and $t_g=\exp[-z \ell^\star_{\rm div}]$. Up to an overall numerical prefactor the transition temperature in the ordered phase is then defined as $t_c=t_g - t_\mu$~\cite{szaboroy:3DTI}. Consequently, $t_c=0$ at the phase boundary between the disordered and ordered (superconducting) phases at zero temperature. As $\ell_{\rm div}$ decreases monotonically with increasing $g_{_0}$, accordingly $t_c$ also increases. Following this prescription, we construct the phase diagram in the $(g_{_0},t)$ plane, where the phase boundary between the disordered (Fermi liquid) and ordered phase (superconductor) marks $t_c$ as a function of the bare strength of $g_{_0}$, as shown in Fig.~\ref{Fig:QAHIPD}. It should, however, be noted that our RG flow equations are operative at any arbitrary temperature and in principle one can arrive at a better estimation of the finite-temperature phase boundaries between the disordered Fermi liquid and spontaneously ordered phases by directly computing the corresponding critical interaction strength $g_c$ for any desired transition temperature $t_c$~\cite{finitemu:3, finitemu:4, finitemu:5}. Such a calculation only quantitatively changes the outcomes in the phase diagrams shown in this work.

%%%%%%%%%%%%%%%%%%%%%%%%%%%%%%%%%%%%%%%%%%%%%%%%%%%%%%%%%%%%%%%%%%%
%%%%%%%%%%%%%%%%%%%%%%%%%%%%%%%%%%%%%%%%%%%%%%%%%%%%%%%%%%%%%%%%%%%
%%%%%%%%%%%%%%%%%%%%%%%%%%%%%%%%%%%%%%%%%%%%%%%%%%%%%%%%%%%%%%%%%%%
%%%%%%%%%%%%%%%%%%%%%%%%%%%%%%%%%%%%%%%%%%%%%%%%%%%%%%%%%%%%%%%%%%%
%%%%%%%%%%%%%%%%%%%%%%%%%%%%%%%%%%%%%%%%%%%%%%%%%%%%%%%%%%%%%%%%%%%
\begin{figure*}[t!]
\includegraphics[width=1.00\linewidth]{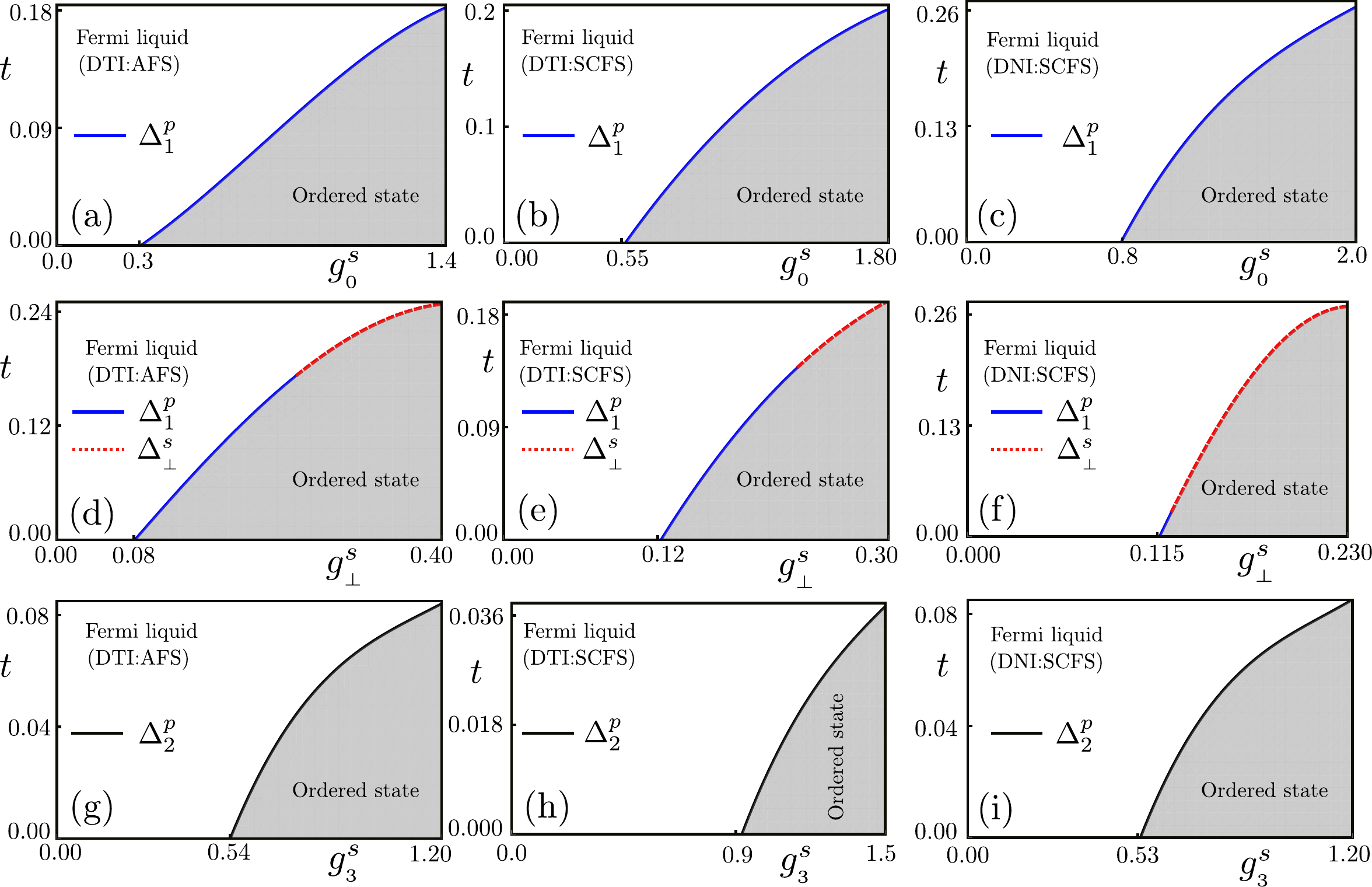}
\caption{Cuts of the global phase diagram for repulsive interactions [(a)-(c)] $g^s_{_0}$, [(d)-(f)] $g_{_\perp}^s$, and [(g)-(i)] $g_{_3}^s$ in the spin-singlet channels (denoted by the superscript $s$) in a time-reversal symmetric doped planar Dirac insulator from class AII [see Eq.~\eqref{eq:LintQSHI}]. The parameter values $(m,b,\mu)$ are equal to (a) $(-0.7,3.2,0.6)$, (b) $(-0.5,3.2,0.6)$, (c) $(0.2,3.2,0.5)$, (d) $(-0.7,3.2,0.6)$, (e) $(-0.5,3.2,0.6)$, (f) $(0.2,3.8,0.3)$, (g) $(-0.9,1.0,0.86)$, (h) $(-0.9,1.0,0.91)$, and (i) $(0.05,0.05,0.86)$. The horizontal (vertical) axis corresponds to dimensionless coupling constant (temperature). The white (shaded) regions correspond to disordered Fermi liquid (ordered state). The nature of the ordered state is color and style (solid or dashed) coded along the phase boundaries between the ordered and disordered phases, which also yield the dimensionless transition temperature ($t_c$). The abbreviations are identical to those in Fig.~\ref{Fig:QAHIPD}. For the physical meaning of each ordered state and its symmetry transformations see Table~\ref{tab:bilinears}. Dimensionless coupling constants ($g^s_{_0}$, $g_{_\perp}^s$, and $g_{_3}^s$) are measured in units of $\epsilon$, where $\epsilon=d-1$.  
}~\label{fig:QSHIPD1}
\end{figure*}
%%%%%%%%%%%%%%%%%%%%%%%%%%%%%%%%%%%%%%%%%%%%%%%%%%%%%%%%%%%%%%%%%%%
%%%%%%%%%%%%%%%%%%%%%%%%%%%%%%%%%%%%%%%%%%%%%%%%%%%%%%%%%%%%%%%%%%%
%%%%%%%%%%%%%%%%%%%%%%%%%%%%%%%%%%%%%%%%%%%%%%%%%%%%%%%%%%%%%%%%%%%
%%%%%%%%%%%%%%%%%%%%%%%%%%%%%%%%%%%%%%%%%%%%%%%%%%%%%%%%%%%%%%%%%%%
%%%%%%%%%%%%%%%%%%%%%%%%%%%%%%%%%%%%%%%%%%%%%%%%%%%%%%%%%%%%%%%%%%%

Notice that the nucleation of the superconducting state from a repulsive density-density interaction in this system follows a selection rule, which we mention here briefly and discuss in details in the next section where the interacting theory is described by multiple local quartic terms and the system fosters a plethora of ordered states including the pairing ones. In the announced Nambu-doubled basis, the matrix appearing with the local four-fermion interaction $\Gamma_{30}$ and two matrices describing the paired state $\Gamma_{13}$ and $\Gamma_{23}$ constitute a O(3) vector composed of three mutually anti-commuting matrices. Therefore, repulsive density-density interaction provides the strongest boost in this pairing channel, by endowing it with the largest positive anomalous dimension, which can be verified explicitly from the Feynman diagrams shown in Fig.~\ref{Fig:FeynDiag_Susceptibility}. The stage is now set to discuss the role of repulsive electronic interactions in a doped two-dimensional Dirac insulator, belonging to class AII (preserving the time-reversal symmetry) in the ten-fold classification scheme.

\section{Doped Dirac Insulator: Class AII}~\label{sec:dopedQSHI}

The Dirac Hamiltonian describing a two-dimensional class AII insulator can be constructed from Eq.~\eqref{eq:DiracUniversal} with $\Gamma_{1}=\sigma_3 \otimes \tau_1$, $\Gamma_{2}=\sigma_3 \otimes \tau_2$, and $\Gamma_{3}=\sigma_3 \otimes \tau_3$. The set of Pauli matrices $\{ \sigma_\nu \}$ operates on the spin indices~\cite{TITSC:8}. The four-component Dirac spinor in this case reads as $\psi^\top_{\vec{k}}=(c_{+,\uparrow}, c_{-,\uparrow}, c_{+,\downarrow}, c_{-,\downarrow})(\vec{k})$, where $c_{\tau,\sigma}(\vec{k})$ is the fermionic annihilation operator with parity eigenvalue $\tau=\pm$, spin projection $\sigma=\uparrow, \downarrow$ in the $z$ direction, and momentum $\vec{k}$.

%%%%%%%%%%%%%%%%%%%%%%%%%%%%%%%%%%%%%%%%%%%%%%%%%%%%%%%%%%%%%%%%%%%
%%%%%%%%%%%%%%%%%%%%%%%%%%%%%%%%%%%%%%%%%%%%%%%%%%%%%%%%%%%%%%%%%%%
%%%%%%%%%%%%%%%%%%%%%%%%%%%%%%%%%%%%%%%%%%%%%%%%%%%%%%%%%%%%%%%%%%%
%%%%%%%%%%%%%%%%%%%%%%%%%%%%%%%%%%%%%%%%%%%%%%%%%%%%%%%%%%%%%%%%%%%
%%%%%%%%%%%%%%%%%%%%%%%%%%%%%%%%%%%%%%%%%%%%%%%%%%%%%%%%%%%%%%%%%%%
\begin{figure*}[t!]
\includegraphics[width=1.00\linewidth]{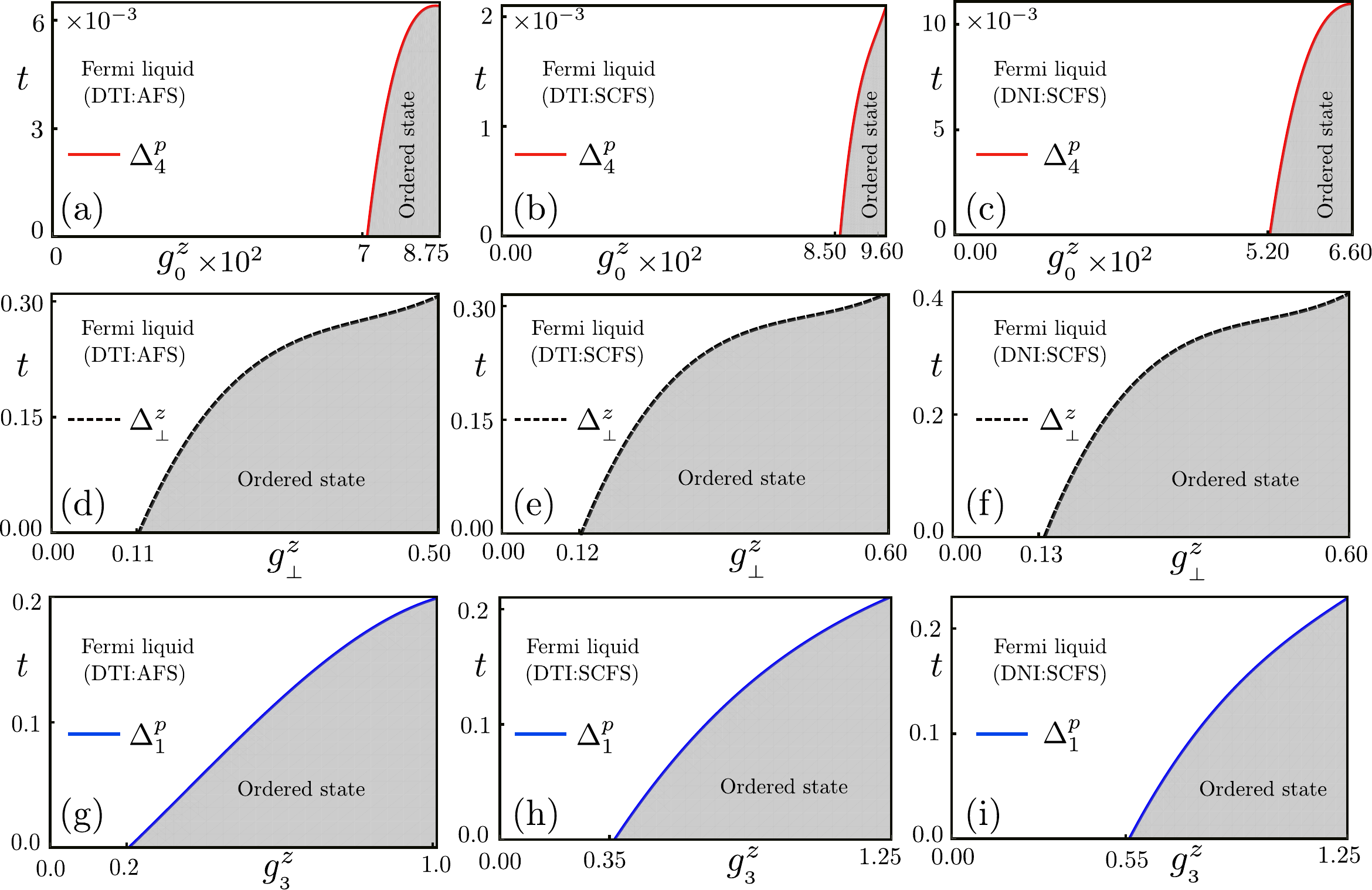}
\caption{Cuts of the global phase diagram for repulsive interactions [(a)-(c)] $g^z_{_0}$, [(d)-(f)] $g_{_\perp}^z$, and [(g)-(i)] $g_{_3}^z$ among fermions with spin projection along the easy-axis or the $z$ direction (denoted by the superscript $z$) in a time-reversal symmetric doped planar class AII Dirac insulator [see Eq.~\eqref{eq:LintQSHI}]. The parameter values $(m,b,\mu)$ are equal to (a) $(-0.7,3.5,0.55)$, (b) $(-0.4,3.5,0.45)$, (c) $(0.1,1.0,0.25)$, (d) $(-0.7,3.5,0.6)$, (e) $(-0.5,3.5,0.6)$, (f) $(0.2,3.5,0.5)$, (g) $(-0.7,3.2,0.6)$, (h) $(-0.5,3.2,0.6)$, and (i) $(0.2,3.2,0.5)$. The rest of the details are the same as in Figs.~\ref{Fig:QAHIPD} and~\ref{fig:QSHIPD1}. For the physical meaning of each ordered state and its symmetry transformations see Table~\ref{tab:bilinears}. Dimensionless coupling constants ($g^z_{_0}$, $g_{_\perp}^z$, and $g_{_3}^z$) are measured in units of $\epsilon=d-1$.
}~\label{fig:QSHIPD2}
\end{figure*}
%%%%%%%%%%%%%%%%%%%%%%%%%%%%%%%%%%%%%%%%%%%%%%%%%%%%%%%%%%%%%%%%%%%
%%%%%%%%%%%%%%%%%%%%%%%%%%%%%%%%%%%%%%%%%%%%%%%%%%%%%%%%%%%%%%%%%%%
%%%%%%%%%%%%%%%%%%%%%%%%%%%%%%%%%%%%%%%%%%%%%%%%%%%%%%%%%%%%%%%%%%%
%%%%%%%%%%%%%%%%%%%%%%%%%%%%%%%%%%%%%%%%%%%%%%%%%%%%%%%%%%%%%%%%%%%
%%%%%%%%%%%%%%%%%%%%%%%%%%%%%%%%%%%%%%%%%%%%%%%%%%%%%%%%%%%%%%%%%%%

 This model preserves the time-reversal symmetry ($\mathcal T$) under which $\mathcal{T} \psi^\ast_{\vec{k}} \mathcal{T}=\Gamma_{21} \psi_{-\vec{k}}$, the parity (${\mathcal P}$) under which $\mathcal{P} \psi_{\vec{k}} \mathcal{P}=\Gamma_{33} \psi_{-\vec{k}}$, and the charge conjugation symmetry (${\mathcal C}$) under which $\mathcal{C} \psi_{\vec{k}} \mathcal{C}=\Gamma_{01} \psi^\ast_{\vec{k}}$, where $\Gamma_{\rho \lambda}=\sigma_\rho \otimes \tau_\lambda$. In addition, the Hamiltonian is invariant under the fourfold rotation about the $z$ direction ($C^z_4$), generated by $R^{z}_{\pi/2}=\exp[-i \Gamma_{03}\pi/4]$ under which $(k_x,k_y) \to (k_y,-k_x)$. The Hamiltonian is also invariant under a fourfold rotation of the spin quantization axis about the $z$ direction ($S^z_4$), generated by $\exp[i\pi\Gamma_{30}/4]$. We impose these symmetries on four-fermion interaction terms taking a generic form $g_{_{\rho\lambda\kappa\beta}}(\psi^\dagger \Gamma_{\rho \lambda}\psi)(\psi^\dagger \Gamma_{\kappa \beta}\psi)$, where a summation over repeated indices ($\rho$, $\lambda$, $\kappa$, and $\beta$) is assumed and $\rho,\lambda,\kappa,\beta=0,1,2,3$.

Once all the above mentioned symmetries are accounted for, the interacting Lagrangian ($L_{\rm int}$) containing all symmetry allowed local four-fermion terms is described by nine coupling constants and is given by 
\allowdisplaybreaks[4]
\begin{eqnarray}~\label{eq:LintQSHI}
L^{\rm AII}_{\rm int} &=& g_{_0}^s (\psi^{\dagger} \Gamma_{00} \psi)^2 + g_{_\perp}^s \sum_{j=1,2}(\psi^{\dagger} \Gamma_{0j}\psi)^2 + g_{_3}^s(\psi^{\dagger} \Gamma_{03}\psi)^2 \nonumber \\
&+& g_{_0}^z (\psi^{\dagger} \Gamma_{30}\psi)^2 + g_{_\perp}^z \sum_{j=1,2} (\psi^{\dagger} \Gamma_{3j}\psi)^2 + g_{_3}^z (\psi^{\dagger} \Gamma_{33}\psi)^2 \nonumber \\
&+& g_{_0}^{_\perp} \sum_{j=1,2} (\psi^{\dagger} \Gamma_{j0}\psi)^2 + g_{_3}^{\perp} \sum_{j=1,2} (\psi^{\dagger} \Gamma_{j3}\psi)^2 \nonumber \\
&+& g_{_\perp}^{\perp} \sum_{j=1,2}[(\psi^{\dagger} \Gamma_{j1}\psi)^2+(\psi^{\dagger} \Gamma_{j2}\psi)^2],
\end{eqnarray} 
where $\psi \equiv \psi_{\vec{k}}$ (for brevity). The subscript in the coupling constants bear the same physical meaning as in Eq.~\eqref{eq:intQAHI}. By contrast, the superscripts therein indicate whether the interaction is in the spin-singlet channel (for ``$s$") or with the spin projection in the easy-axis or $z$ direction (for ``$z$") or among fermions with spin confined in the easy-plane or $xy$ plane (for ``$\perp$"). However, only four quartic terms are linearly independent due to the Fierz relations among nine four-fermions terms, as shown in Appendix~\ref{append:fierz}~\cite{Fierz:1, Fierz:2, Fierz:3, Fierz:4, Fierz:5}. Without any loss of generality, we choose four linearly independent quartic terms to be the ones accompanying the coupling constants $g_{_0}^s$, $g_{_\perp}^s$, $g_{_3}^s$, and $g_{_3}^z$. The imaginary time action for this system containing the free-fermion part ($S_0$) and the interacting part ($S_{\rm int}$) assumes the form shown in Eq.~\eqref{eq:actionInt}. Now the Dirac Hamiltonian entering $S_0$ is expressed in terms of four-dimensional Hermitian $\Gamma$ matrices, mentioned above, and $S_{\rm int}$ is expressed in terms of the chosen four quartic terms.

In order to capture all possible particle-hole and particle-particle orders within a unified framework, we next introduced the following Nambu-doubled spinor basis $\psi^\top_{\rm Nam}=(\psi_{\omega_n,k}, \Gamma_{21}\psi_{-\omega_n,-k}^\star)$. In this basis, the massive Dirac Hamiltonian takes the form
\begin{equation}~\label{HDirNamAII}
\hat{h}^{\rm Nam}_{\rm Dir, AII}=v \left( \Gamma_{331} k_x + \Gamma_{332} k_y \right) + (m+b k^2) \Gamma_{333}-\mu \Gamma_{300},
\end{equation}
where the eight-dimensional Hermitian $\Gamma$ matrices are defined as $\Gamma_{\alpha \nu \rho}=\eta_\alpha \otimes \sigma_\nu \otimes \tau_\rho$. Accordingly, the fermionic Green's function takes the form shown in Eq.~\eqref{eq:GreensNambu}, with 
\begin{equation}
G_\pm = \frac{\Omega_\pm \Gamma_{00} \pm v (\Gamma_{31} k_x + \Gamma_{32} k_y) \pm (m+b k^2) \Gamma_{33}}{\Omega^2_\pm-E^2_k}.
\end{equation}
In this Nambu-doubled basis the matrices appearing with four chosen quartic interaction terms transform according to $\Gamma_{00} \to \Gamma_{300}$ in $g_{_0}^s$, $\Gamma_{01} \to \Gamma_{301}$ and $\Gamma_{02} \to \Gamma_{302}$ in $g_{_\perp}^s$, $\Gamma_{03} \to \Gamma_{003}$ in $g_{_3}^s$, and $\Gamma_{33} \to \Gamma_{333}$ in $g_{_3}^z$. The imaginary time action containing all the local (momentum-independent) source terms takes the form shown in Eq.~\eqref{eq:actionSource}, with
\allowdisplaybreaks[4]
\begin{eqnarray}~\label{eq:excQSHI}
&&h^{\rm AII}_{\rm exc} =\nonumber \\
&& \Delta_0^s (\psi^{\dagger} \Gamma_{300}\psi) + \Delta_{\perp}^s \sum_{j=1,2} (\psi^{\dagger} \Gamma_{30j}\psi) + \Delta_3^s (\psi^{\dagger} \Gamma_{003}\psi) \nonumber \\
&+& \Delta_0^z (\psi^{\dagger} \Gamma_{030} \psi) +  \Delta_{\perp}^z \sum_{j=1,2} (\psi^{\dagger} \Gamma_{03j}\psi) + \Delta_3^z(\psi^{\dagger} \Gamma_{333}\psi)
 \nonumber \\
&+& \Delta_0^{\perp} \sum_{j=1,2} (\psi^{\dagger} \Gamma_{0j0}\psi) + \Delta_3^{\perp}\sum_{j=1,2} (\psi^{\dagger} \Gamma_{3j3}\psi)  \nonumber \\
&+& \Delta_{\perp}^{\perp} \sum_{j=1,2} [(\psi^{\dagger} \Gamma_{0j1}\psi)+(\psi^{\dagger} \Gamma_{0j2}\psi)]
\end{eqnarray}
and 
\allowdisplaybreaks[4]
\begin{eqnarray}~\label{eq:pairQSHI}
h^{\rm AII}_{\rm pair} &=& \sum_{\alpha=1,2} \bigg[\Delta_1^p (\psi^\dagger \Gamma_{\alpha 00}\psi) + \Delta_2^p \sum_{j=1,2}(\psi^\dagger \Gamma_{\alpha 0j}\psi) \nonumber \\
&+& \Delta_3^p (\psi^\dagger \Gamma_{\alpha 33}\psi)+\Delta_4^p \sum_{j=1,2} (\psi^\dagger \Gamma_{\alpha j3}\psi) \bigg],
\end{eqnarray}  
where $\psi \equiv \psi_{\rm Nam}$, and $\alpha$ taking values 1 and 2, reflects the U(1) gauge redundancy in defining the superconducting phase. Transformations of all the fermion bilinears under various symmetries of the noninteracting system are summarized in Table~\ref{tab:bilinears}, where we also mention their physical meaning or nature. 

%%%%%%%%%%%%%%%%%%%%%%%%%%%%%%%%%%%%%%%%%%%%%%%%%%%%%%%%%%%%%%%%%%%
%%%%%%%%%%%%%%%%%%%%%%%%%%%%%%%%%%%%%%%%%%%%%%%%%%%%%%%%%%%%%%%%%%%
%%%%%%%%%%%%%%%%%%%%%%%%%%%%%%%%%%%%%%%%%%%%%%%%%%%%%%%%%%%%%%%%%%%
%%%%%%%%%%%%%%%%%%%%%%%%%%%%%%%%%%%%%%%%%%%%%%%%%%%%%%%%%%%%%%%%%%%
%%%%%%%%%%%%%%%%%%%%%%%%%%%%%%%%%%%%%%%%%%%%%%%%%%%%%%%%%%%%%%%%%%%
\begin{figure*}[t!]
\includegraphics[width=1.00\linewidth]{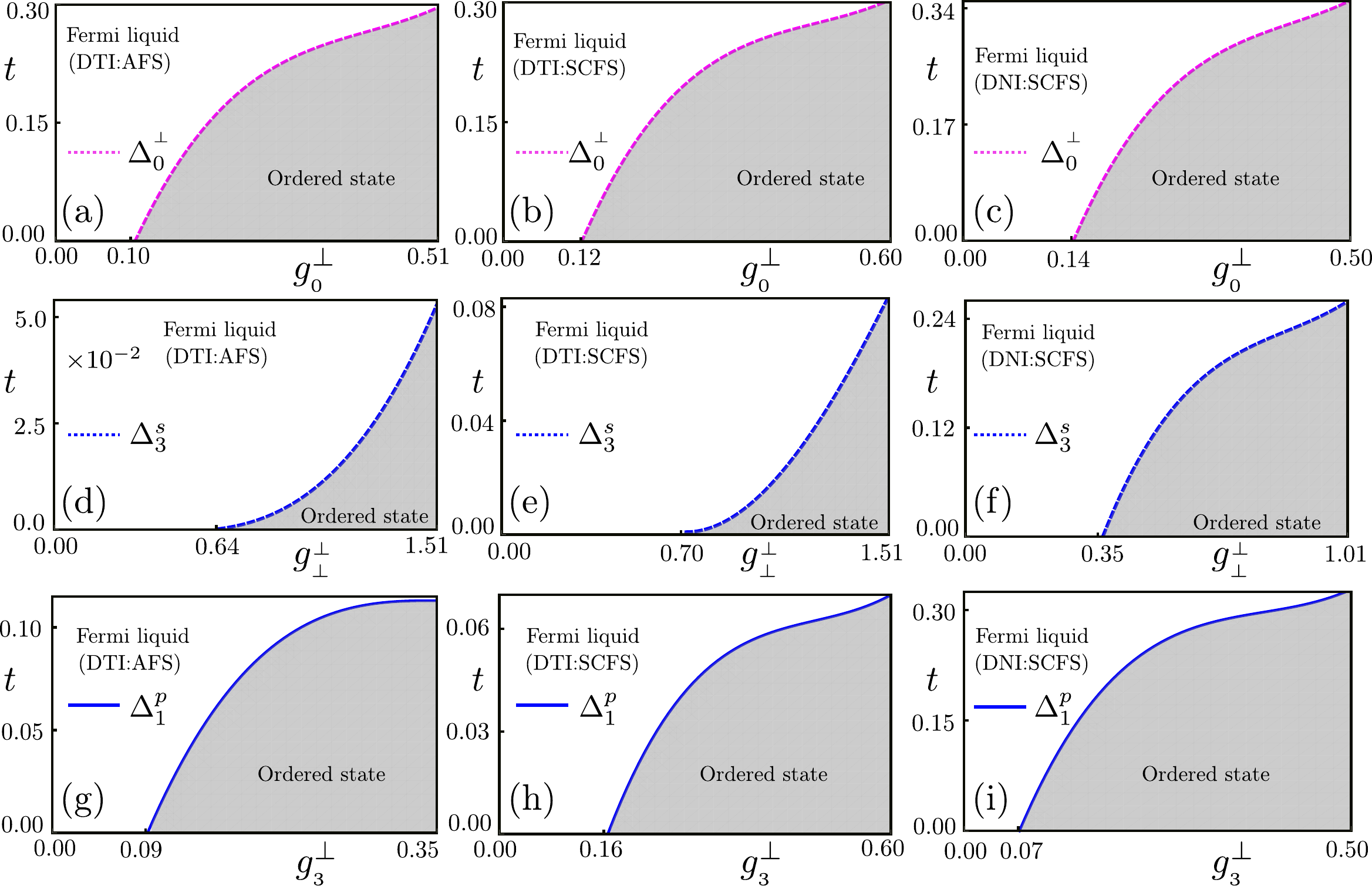}
\caption{Cuts of the global phase diagram for repulsive interactions [(a)-(c)] $g^\perp_{_0}$, [(d)-(f)] $g_{_\perp}^\perp$, and [(g)-(i)] $g_{_3}^\perp$ among fermions with spin projection confined within the easy or $xy$ plane (denoted by the superscript $\perp$) in a time-reversal symmetric doped planar class AII Dirac insulator [see Eq.~\eqref{eq:LintQSHI}]. The parameter values $(m,b,\mu)$ are equal to (a) $(-0.7,3.5,0.6)$, (b) $(-0.5,3.5,0.6)$, (c) $(0.2,3.5,0.5)$, (d) $(-0.7,2.7,0.66)$, (e) $(-0.5,2.6,0.6)$, (f) $(0.2,1.0,0.6)$, (g) $(-0.9,1.0,0.86)$, (h) $(-0.9,1.0,0.91)$, and (i) $(0.2,1.0,0.6)$. The rest of the details are the same as in Figs.~\ref{Fig:QAHIPD} and~\ref{fig:QSHIPD1}. For the physical meaning of each ordered state and its symmetry transformations see Table~\ref{tab:bilinears}. Dimensionless coupling constants ($g^\perp_{_0}$, $g_{_\perp}^\perp$, and $g_{_3}^\perp$) are measured in units of $\epsilon=d-1$.
}~\label{fig:QSHIPD3}
\end{figure*}
%%%%%%%%%%%%%%%%%%%%%%%%%%%%%%%%%%%%%%%%%%%%%%%%%%%%%%%%%%%%%%%%%%%
%%%%%%%%%%%%%%%%%%%%%%%%%%%%%%%%%%%%%%%%%%%%%%%%%%%%%%%%%%%%%%%%%%%
%%%%%%%%%%%%%%%%%%%%%%%%%%%%%%%%%%%%%%%%%%%%%%%%%%%%%%%%%%%%%%%%%%%
%%%%%%%%%%%%%%%%%%%%%%%%%%%%%%%%%%%%%%%%%%%%%%%%%%%%%%%%%%%%%%%%%%%
%%%%%%%%%%%%%%%%%%%%%%%%%%%%%%%%%%%%%%%%%%%%%%%%%%%%%%%%%%%%%%%%%%%

In Appendix~\ref{append:bandprojection}, we show that close to the Fermi surface, $\Delta^p_1$ and $\Delta^p_3$ paired states assume the form of a spin-singlet $s$-wave superconductor with the amplitude of the gap $\sim m$ (Dirac mass) in the latter one. Thus, $\Delta^p_3$ pairing yields a gapless Fermi surface of neutral Majorana fermions in doped Dirac semimetal with $m=0$ as found at the quantum critical point separating two topologically distinct insulators (see Fig.~\ref{fig:FStopology}). Close to the Fermi surface, the fully gapped $\Delta^p_2$ paired state represents a rotational symmetry breaking class C nematic superconductor, with the effective reduced BCS Hamiltonian taking the form of two identical copies of the time-reversal symmetry breaking $p+ip$ pairing. Finally, we note that the $\Delta^p_4$ paired state represents an isotropic and time-reversal symmetric fully gapped $p \pm ip$ superconductor around the Fermi surface, a prototypical representative of class DIII superconductors. It should be noted that the degeneracy of the $p \pm ip$ pairing resulting from a single local pairing is protected by the time-reversal symmetry in the normal state of a class AII system. If we sacrifice the fourfold rotational symmetry in the normal state that gets inherited by the paired state as well, which then represents a nematic $p \pm ip$ pairing with identical topological properties as the isotropic one as this paired state transforms as a scalar under fourfold rotations (see Table~\ref{tab:bilinears}).

\begin{widetext}
The coarse graining procedure to arrive at the RG flow equations for the quartic interaction and source terms has already been discussed in details in the previous section. So, here we only quote the results. The RG flow equations for four chosen quartic interaction terms in terms of dimensionless coupling constants defined as $2 \pi g^s_{_0} \Lambda^\epsilon/v \to g^s_{_0}$, $2 \pi g_{_\perp}^s \Lambda^\epsilon/v \to g_{_\perp}^s$, $2 \pi g_{_3}^s \Lambda^\epsilon/v \to g_{_3}^s$, and $2 \pi g_3^z \Lambda^\epsilon/v \to g_3^z$, obtained after computing the Feynman diagrams shown in Fig.~\ref{Fig:FeynDiag_Interaction}, are explicitly given by (see Appendix~\ref{append:Feynman} for details)
\allowdisplaybreaks[4]
\begin{eqnarray}~\label{eq:RGcouplingAII}
\frac{dg_{_0}^s}{d\ell} &=&- \epsilon g_{_0}^s + (-8 f_k  - 4 f_0 - 2 f_w- 2 \tilde f_w) (g^{s}_{_0})^2
+ (8 f_k  +  8 \tilde f_k  + 8 f_0  + 8 f_w) g^s_{_0} g^s_{_\perp} 
+(- 4 f_0  +  4 \tilde f_0  + 4 f_w  - 4 \tilde f_w)( g^s_{_\perp})^2 \nonumber \\
&+& (8 f_k  - 4 \tilde f_0  + 4 f_w) g^s_{_0} g^s_{_3} + ( - 8 f_k  - 8 \tilde f_k) g^s_{_\perp} g^s_{_3} 
+ (2 f_w - 2\tilde f_w) (g_{_3}^s)^2 + ( 8 f_k + 8 f_0  + 4 \tilde f_0  + 4 f_w) g_{_0}^s g_{_3}^z \nonumber \\
&+& ( 8 f_k  + 8 \tilde f_k) g_{_\perp}^s g_{_3}^z + (-4 f_w + 4 \tilde f_w) g_{_3}^s g_{_3}^z + (2 f_w - 2 \tilde f_w) (g_{_3}^z)^2,
\nonumber \\
%%%%%%%%%%%%%%%%%%%%%%%%%%%%%%%%%%%%%%%%%%%%%%%%%%%%%%%%%%%%%%%%%%%%%%%%%%%%%%%%%%%%%%%%%%%%%%%%%%%%%%%%%%%%%%%%%%%%%
\frac{d g_{_\perp}^s}{d \ell} &=& -\epsilon g_{_\perp}^s + (-2 f_k  + 2 \tilde f_k) (g_{_0}^s)^2+( - 8 f_0  + 4 \tilde f_0  + 8 f_w - 
 4 \tilde f_w) g_{_0}^s g^s_{_\perp}+( - 8 f_k  + 8\tilde f_k  + 8 f_0 - 8 f_w) (g^s_{_\perp})^2  \nonumber \\
&+& (- 4 f_k - 4 \tilde f_k) g_{_0}^s g_{_3}^s + ( - 4 \tilde f_0 + 4 \tilde f_w) g_{_\perp}^s g_{_3}^s
+( -2 f_k  + 2 \tilde f_k )(g_{_3}^s)^2 +( 4 f_k + 4\tilde f_k ) g_{_0}^s g_{_3}^z  \nonumber \\
&+& ( 8 f_0 +  4 \tilde f_0  - 8 f_w  - 4 \tilde f_w ) g_{_\perp}^s g_{_3}^z 
+ (4 f_k  -  4 \tilde f_k) g_{_3}^s g_{_3}^z + (- 2 f_k + 2 \tilde f_k) (g_{_3}^z)^2,
\nonumber \\
%%%%%%%%%%%%%%%%%%%%%%%%%%%%%%%%%%%%%%%%%%%%%%%%%%%%%%%%%%%%%%%%%%%%%%%%%%%%%%%%%%%%%%%%%%%%%%%%%%%%%%%%%%%%%%%%%%%%%
\frac{dg_{_3}^s}{d \ell} &=&-\epsilon g_{_3}^s + (4 f_k  - 4\tilde f_k) (g_{_0}^s)^2
+( - 8 f_k  + 8 \tilde f_k   + 8 f_0 -  8 \tilde f_0) g_{_0}^s g_{_\perp}^s 
+ (16 f_k  - 16 \tilde f_k - 4 f_0  + 4 \tilde f_0 +  4 f_w + 4 \tilde f_w) (g_{_\perp}^s)^2  \nonumber \\
&+& ( 8 \tilde f_k - 4 \tilde f_0  +  8 f_w - 4 \tilde f_w) g_{_0}^s g_{_3}^s 
+( 8 f_k  - 8 \tilde f_k  +  8 \tilde f_0 - 8 f_w) g_{_\perp}^s g_{_3}^s 
+ (12 f_k  - 4 \tilde f_k  - 4 f_0 - 4 f_w) (g_{_3}^s)^2 \nonumber \\
&+& (8 f_k - 8 \tilde f_k  + 8 f_w  + 8 \tilde f_w) g_{_\perp}^s g_{_3}^z 
+(- 8 f_k + 8 f_0  + 4 \tilde f_0  +  4 \tilde f_w) g_{_3}^s g_{_3}^z + (4 f_k - 4 \tilde f_k) (g_{_3}^z)^2,
\nonumber
\end{eqnarray}
%%%%%%%%%%%%%%%%%%%%%%%%%%%%%%%%%%%%%%%%%%%%%%%%%%%%%%%%%%%%%%%%%%%%%%%%%%%%%%%%%%%%%%%%%%%%%%%%%%%%%%%%%%%%%%%%%%%%%
\begin{eqnarray}
\text{and} \:\:\:
\frac{dg_3^z}{d \ell} &=& -\epsilon g_{_3}^z + (4 f_k - 4 \tilde f_k  + 2 f_0 + 2 \tilde f_0) (g_{_0}^s)^2 
+ (16\tilde f_k +  8 f_0  - 8 \tilde f_0) g_{_0}^s g_{_\perp}^s 
+ (16 f_k  - 16 \tilde f_k +  8 \tilde f_0) (g_{_\perp}^s)^2  \nonumber \\
&+& (8 f_k  + 8 \tilde f_k  - 4 f_0 -  4 \tilde f_0) g_{_0}^s g_{_3}^s 
+ ( - 16 \tilde f_k  + 8 f_0 + 8 \tilde f_0) g_{_\perp}^s g_{_3}^s 
+ (4 f_k - 4 \tilde f_k + 2 f_0 + 2 \tilde f_0) (g_{_3}^s)^2 \nonumber \\
&+& ( - 8 f_k  +  4 f_0  + 8 f_w  - 4 \tilde f_w) g_{_0}^s g_{_3}^z 
+ (16 f_k -  8 f_0 + 8 \tilde f_w) g_{_\perp}^s g_{_3}^z 
+( - 8 f_k  + 4 f_0  + 4 \tilde f_w) g_{_3}^s g_{_3}^z \nonumber \\
&+& (12 f_k  - 4\tilde f_k  - 2 f_0 +  2 \tilde f_0  - 4 f_w) (g_{_3}^z)^2.
\end{eqnarray}
Various functions appearing in the above equations have already been defined in Eq.~\eqref{eq:flowfunctions}. The RG flow equations for various dimensionless parameters, namely $t$, $\mu$, $m$, and $b$, are given in Eq.~\eqref{eq:Parameters}.

On the other hand, the leading-order RG flow equations of all the source terms in the particle-hole or excitonic channels resulting from the Feynman diagrams shown in Fig.~\ref{Fig:FeynDiag_Susceptibility} in terms of the dimensionless coupling constants defined above are explicitly given by (see Appendix~\ref{append:Feynman} for details)
\allowdisplaybreaks[4]
\begin{eqnarray}~\label{eq:RGsourcePHAII}
\bar{\beta}_{\Delta_0^s} &=& \frac{d \ln \Delta_0^s}{d \ell} -z= -(12 f_k  + 6 f_0  + 6 f_w) g_{_0}^s + (8 f_k  + 4 f_0  + 4 f_w) g_{_\perp}^s +  (4 f_k  + 2 f_0  + 2 f_w) g_{_3}^s +( 4 f_k + 2 f_0  + 2 f_w) g_{_3}^z, \nonumber \\
%%%%%%%%%%%%%%%%%%%%%%%%%%%%%%%%%%%%%%%%%%%%%%%%%%%%%%%%%%%%%%%%%%%%%%%%%%%%%%%%%%%%%%%%%%%%%%%%%%%%%%%%%%%%%%%%%%%%%
\bar{\beta}_{\Delta_{\perp}^s} &=& \frac{d \ln \Delta_{\perp}^s}{d \ell} -z= -(2 f_0  - 2 f_w) g_{_0}^s +( 8 f_0  - 8 f_w) g_{_\perp}^s + (2 f_0  - 2 f_w) g_{_3}^s +  (2 f_0 - 2 f_w) g_{_3}^z, \nonumber \\
%%%%%%%%%%%%%%%%%%%%%%%%%%%%%%%%%%%%%%%%%%%%%%%%%%%%%%%%%%%%%%%%%%%%%%%%%%%%%%%%%%%%%%%%%%%%%%%%%%%%%%%%%%%%%%%%%%%%%
\bar{\beta}_{\Delta_3^s} &=& \frac{d \ln \Delta_3^s}{d \ell} -z= -(4 f_k  - 2 f_0 - 2 f_w) g_{_0}^s +( 8 f_k  - 4 f_0  - 4 f_w) g_{_\perp}^s + ( 12 f_k  - 6 f_0  - 6 f_w) g_{_3}^s - (4 f_k  - 2 f_0  - 2 f_w) g_{_3}^z, \nonumber \\
%%%%%%%%%%%%%%%%%%%%%%%%%%%%%%%%%%%%%%%%%%%%%%%%%%%%%%%%%%%%%%%%%%%%%%%%%%%%%%%%%%%%%%%%%%%%%%%%%%%%%%%%%%%%%%%%%%%%%
\bar{\beta}_{\Delta_0^z} &=& \frac{d \ln \Delta_0^z}{d \ell} -z= (4 f_k  + 2 f_k  + 2 f_w) g_{_0}^s +( 8 f_k  + 4 f_0  + 4 f_w) g_{_\perp}^s +  (4 f_k  + 2 f_0  + 2 f_w) g_{_3}^s + (4 f_k  + 2 f_0  + 2 f_w) g_{_3}^z, \nonumber \\
%%%%%%%%%%%%%%%%%%%%%%%%%%%%%%%%%%%%%%%%%%%%%%%%%%%%%%%%%%%%%%%%%%%%%%%%%%%%%%%%%%%%%%%%%%%%%%%%%%%%%%%%%%%%%%%%%%%%%
\bar{\beta}_{\Delta_{\perp}^z} &=& \frac{d \ln \Delta_{\perp}^z}{d \ell} -z= -(2 f_0  - 2 f_w) g_{_0}^s + (2 f_0  - 2 f_w) g_{_3}^s 
+( 2 f_0  - 2 f_w) g_{_3}^z, \nonumber \\
%%%%%%%%%%%%%%%%%%%%%%%%%%%%%%%%%%%%%%%%%%%%%%%%%%%%%%%%%%%%%%%%%%%%%%%%%%%%%%%%%%%%%%%%%%%%%%%%%%%%%%%%%%%%%%%%%%%%%
\bar{\beta}_{\Delta_3^z} &=& \frac{d \ln \Delta_3^z}{d \ell} -z= -(4 f_k  - 2 f_0  - 2 f_w) g_{_0}^s + (8 f_k  - 4 f_0  - 4 f_w) g_{_\perp}^s 
- ( 4 f_k  - 2 f_0  - 2 f_w) g_{_3}^s + (12 f_k  - 6 f_0  - 6 f_w) g_{_3}^z, \nonumber \\
%%%%%%%%%%%%%%%%%%%%%%%%%%%%%%%%%%%%%%%%%%%%%%%%%%%%%%%%%%%%%%%%%%%%%%%%%%%%%%%%%%%%%%%%%%%%%%%%%%%%%%%%%%%%%%%%%%%%%
\bar{\beta}_{\Delta_0^\perp} &=& \frac{d \ln \Delta_0^\perp}{d \ell} -z= -(4 f_k  + 2 f_0  - 2 f_w) g_{_0}^s - ( 8 f_k  + 4 f_0  - 4 f_w) g_{_\perp}^s - ( 4 f_k  + 2 f_0  - 2 f_w) g_{_3}^s + (4 f_k  + 2 f_0  - 2 f_w) g_{_3}^z, \nonumber \\
%%%%%%%%%%%%%%%%%%%%%%%%%%%%%%%%%%%%%%%%%%%%%%%%%%%%%%%%%%%%%%%%%%%%%%%%%%%%%%%%%%%%%%%%%%%%%%%%%%%%%%%%%%%%%%%%%%%%%
\bar{\beta}_{\Delta_\perp^\perp} &=& \frac{d \ln \Delta_\perp^\perp}{d \ell} -z= (2 f_0  + 2 f_w) g_{_0}^s +(- 2 f_0 - 2 f_w) g_{_3}^s + (2 f_0  + 2 f_w) g_{_3}^z, \nonumber \\
%%%%%%%%%%%%%%%%%%%%%%%%%%%%%%%%%%%%%%%%%%%%%%%%%%%%%%%%%%%%%%%%%%%%%%%%%%%%%%%%%%%%%%%%%%%%%%%%%%%%%%%%%%%%%%%%%%%%%
\text{and} \nonumber \\
\bar{\beta}_{\Delta_3^\perp} &=& \frac{d \ln \Delta_3^\perp}{d \ell} -z= (4 f_k  - 2 f_0  + 2 f_w) g_{_0}^s - ( 8 f_k  - 4 f_0  + 4 f_w) g_{_\perp}^s + (4 f_k  - 2 f_0  + 2 f_w) g_{_3}^s - ( 4 f_k  - 2 f_0  + 2 f_w) g_{_3}^z. 
%\nonumber \\
\end{eqnarray}
Finally, the RG flow equations for the source terms in the particle-particle or superconducting channels are given by 
\allowdisplaybreaks[4]
\begin{eqnarray}~\label{eq:RGsourcePPAII}
\bar{\beta}_{\Delta_1^p} &=& \frac{d \ln \Delta_1^p}{d \ell} = (4 \tilde f_k  + 2 \tilde f_0  - 2 \tilde f_w) g_{_0}^s 
+ ( 8 \tilde f_k  + 4\tilde f_0  -  4 \tilde f_w) g_{_\perp}^s - (4\tilde f_k  + 2 \tilde f_0  - 2 \tilde f_w) g_{_3}^s 
+( 4 \tilde f_k  +  2 \tilde f_0 - 2 \tilde f_w) g_{_3}^z, \nonumber \\
%%%%%%%%%%%%%%%%%%%%%%%%%%%%%%%%%%%%%%%%%%%%%%%%%%%%%%%%%%%%%%%%%%%%%%%%%%%%%%%%%%%%%%%%%%%%%%%%%%%%%%%%%%%%%%%%%%%%%
\bar{\beta}_{\Delta_2^p} &=& \frac{d \ln \Delta_2^p}{d \ell}= -(2 \tilde f_0  + 2 \tilde f_w) g_{_0}^s - (2 \tilde f_0 + 2 \tilde f_w) g_{_3}^s +( 2 \tilde f_0  +  2 \tilde f_w) g_{_3}^z, \nonumber \\
%%%%%%%%%%%%%%%%%%%%%%%%%%%%%%%%%%%%%%%%%%%%%%%%%%%%%%%%%%%%%%%%%%%%%%%%%%%%%%%%%%%%%%%%%%%%%%%%%%%%%%%%%%%%%%%%%%%%%
\bar{\beta}_{\Delta_3^p} &=& \frac{d \ln \Delta_3^p}{d \ell} = -(4 \tilde f_k  - 2 \tilde f_0  + 2 \tilde f_w) g_{_0}^s 
+( 8 \tilde f_k  - 4 \tilde f_0  +  4 \tilde f_w) g_{_\perp}^s + (4 \tilde f_k  - 2 \tilde f_0  + 2 \tilde f_w) g_{_3}^s 
-(4 \tilde f_k -  2 \tilde f_0  + 2 \tilde f_w) g_{_3}^z, \nonumber \\
%%%%%%%%%%%%%%%%%%%%%%%%%%%%%%%%%%%%%%%%%%%%%%%%%%%%%%%%%%%%%%%%%%%%%%%%%%%%%%%%%%%%%%%%%%%%%%%%%%%%%%%%%%%%%%%%%%%%%
\text{and} \nonumber \\
\bar{\beta}_{\Delta_4^p} &=& \frac{d \ln \Delta_4^p}{d \ell} = (4 \tilde f_k  - 2 \tilde f_0 - 2 \tilde f_w) g_{_0}^s
-( 8 \tilde f_k  - 4 \tilde f_0  -  4 \tilde f_w) g_{_\perp}^s - ( 4 \tilde f_k  - 2 \tilde f_0  - 2 \tilde f_w) g_{_3}^s
-( 4\tilde  f_k  - 2 \tilde f_0  - 2 \tilde f_w) g_{_3}^z. 
\end{eqnarray}
\end{widetext}
The construction of various cuts of the global phase diagram of an interacting doped Dirac insulator belonging to class AII can be obtained by slightly generalizing the prescription from the previous section. To this end, we numerically solve the coupled RG flow equations from Eqs.~\eqref{eq:RGcouplingAII}-\eqref{eq:RGsourcePPAII} simultaneously.

For any initial condition of the coupling constants we simultaneous run the RG flow equations for all the quartic terms and the source terms, which we stop at an RG scale $\ell^\star_\mu$. For weak enough strength of any interaction none of the quartic couplings diverge at $\ell=\ell^\star_\mu$ and the phase corresponds to a Fermi liquid without any ordering. Nevertheless, beyond a critical strength of the interaction at least one of the coupling constants diverges at this scale, indicating the onset of an ordered state. The nature of the ordered state is then determined from the RG flow of the source terms. Namely, at the RG scale of $\ell=\ell^\star_\mu$, whichever source term receives the largest positive contribution the corresponding fermion bilinear acquires a finite vacuum expectation value, yielding a spontaneous symmetry brekaing. Primarily, we follow this approach to construct the phase diagram for each individual quartic interaction appearing in $L_{\rm int}$ while setting all the remaining couplings to zero at the bare level. Although, we perform the whole analysis at sufficiently low bare temperature $t(0)=5 \times 10^{-6}$, following the prescription from the previous section, we construct various cuts of the phase diagram in the plane of interaction strength and temperature. The results are shown in Figs.~\ref{fig:QSHIPD1}-\ref{fig:QSHIPD3}. Due to the presence of an underlying Fermi surface at low temperature and weak repulsive interactions (still stronger than the critical ones), the system typically fosters various superconducting orders, while for some interaction channels we find the appearance of an excitonic state at higher temperature and stronger repulsive coupling. These observation can be supported from a set of selection rules (SRs) and an organizing principle~\cite{finitemu:3, finitemu:4, finitemu:5}, which we discuss next.

\subsection{Selection rules and organizing principle}~\label{subsec:SROP}

For the sake of simplicity, here we consider a single quartic term $g_{_Q} \sum_i (\psi^\dag_{\rm Nam} Q_i \psi_{\rm Nam})^2$ and a fermion bilinear $\Delta_{R}\sum_i \psi^\dag_{\rm Nam} R_i \psi_{\rm Nam}$, both expressed in the Nambu-doubled basis. Respectively, they are chosen from the sets of all the symmetry-allowed four-fermion interaction terms [Eq.~\eqref{eq:LintQSHI}] and the order parameter or the source fields [Eqs.~\eqref{eq:excQSHI} and~\eqref{eq:pairQSHI}]. Here $Q_i$ and $R_i$ are Hermitian matrices. For concreteness, all the interactions are considered to be repulsive ($g_{_Q}>0$). We denote the number of anticommuting matrix pairs between the four-fermion interaction ($\vec{Q}$) and order parameter ($\vec{R}$) terms as $A_M$. Then from the set of all available ordered phases, the repulsive four-fermion interaction $g_{_Q}$ maximally boosts the nucleation of the ones for which $\vec{Q} \equiv \vec{R}$ (SR1) or $A_M$ is maximal (SR2). As shown in Table~\ref{tab:individual_channels}, all the emergent superconducting states at low temperature resulting from repulsive electronic interactions follow SR2.

%%%%%%%%%%%%%%%%%%%%%%%%%%%%%%%%%%%%%%%%%%%%%%%%%%%%%%%%%%%%%%%%%%%%%%%%%%%%%%%
%%%%%%%%%%%%%%%%%%%%%%%%%%%%%%%%%%%%%%%%%%%%%%%%%%%%%%%%%%%%%%%%%%%%%%%%%%%%%%%
%%%%%%%%%%%%%%%%%%%%%%%%%%%%%%%%%%%%%%%%%%%%%%%%%%%%%%%%%%%%%%%%%%%%%%%%%%%%%%%
%%%%%%%%%%%%%%%%%%%%%%%%%%%%%%%%%%%%%%%%%%%%%%%%%%%%%%%%%%%%%%%%%%%%%%%%%%%%%%%
%%%%%%%%%%%%%%%%%%%%%%%%%%%%%%%%%%%%%%%%%%%%%%%%%%%%%%%%%%%%%%%%%%%%%%%%%%%%%%%
\renewcommand*{\arraystretch}{1.3}
\begin{table}[t!]
    \centering
    \begin{tabular}{|c|c |c|c| c|c | c|c|}
    \hline
      \multirow{2}{*}{CC} &  \multicolumn{3}{c|}{Excitonic order} & \multicolumn{3}{c|}{Pairing order} & ES \\
      \cline{2-7}
            & CF & Symmetry & SR & CF & Symmetry & SR & (SR3)  \\
      \hline
      $g_{_0}^s$ &  -- &  -- & -- & $\Delta^{\rm p}_1$ & O(2) & SR2 &-- \\
      $g_{_\perp}^s$ & $\Delta^{\rm s}_{\perp}$ & O(2) & SR1 & $\Delta^{\rm p}_1$ & O(2) & SR2 & O(4) \\
      $g_{_3}^s$ & $\mathbf{\Delta^{\rm s}_3}$ & $\mathbf{Z_2}$ &\textbf{ SR1} & $\Delta^{\rm p}_2$ & O(2) & SR2 & O(3) \\
      $g_{_0}^z$ & $\mathbf{\Delta^{\rm \perp}_0}$ & \textbf{O(2)} & \textbf{SR2} & $\Delta^{\rm p}_4$ & O(2) & SR2 & O(3) \\
      $g_{_\perp}^z$ & $\Delta^{ z}_{\perp}$ & O(2) & SR1 &-- & -- & -- & -- \\
      $g_{_3}^z$ &-- &-- & -- & $\Delta^{\rm p}_1$ & O(2) & SR2 &-- \\
      $g_{_0}^{\perp}$ & $\Delta^{\rm \perp}_0$    & O(2) & SR2 & $-$ & -- & -- & -- \\
      $g_{_\perp}^{\perp}$ & $\Delta^{\rm s}_3$ & $Z_2$ & SR2 & --&-- & -- & -- \\
      $g_{_3}^{\perp}$ &  -- &  -- & -- & $\Delta^{\rm p}_0$ & O(2) & SR2 &-- \\
      \hline
    \end{tabular}
    \caption{Excitonic and pairing instabilities, indicated by their respective conjugate fields (CFs), for individual repulsive interactions with a single specific coupling constant (CC), with all other CCs set to zero. We highlight the role of the selection rule (SR) for each such order and the symmetry of the individual order parameter. The right most column indicates the enlarged symmetry (ES) of the composite order parameter, constructed by combining those for the excitonic and pairing orders, following SR3. Dashed lines indicate the absence of any excitonic or pairing order for the corresponding repulsive interaction and hence any ES of the composite order parameter. Excitonic orders shown in bold font do not appear in the corresponding phase diagram within the range of repulsive interaction shown therein. They appear for even stronger repulsive interactions, which is not shown explicitly in the corresponding phase diagrams, displayed in Figs.~\ref{fig:QSHIPD1}-\ref{fig:QSHIPD3}. For the discussion on the SRs see Sec.~\ref{subsec:SROP}. An ordered state with $Z_2$ symmetry is described by a single matrix operator.  
		}~\label{tab:individual_channels}
\end{table}
%%%%%%%%%%%%%%%%%%%%%%%%%%%%%%%%%%%%%%%%%%%%%%%%%%%%%%%%%%%%%%%%%%%%%%%%%%%%%%%
%%%%%%%%%%%%%%%%%%%%%%%%%%%%%%%%%%%%%%%%%%%%%%%%%%%%%%%%%%%%%%%%%%%%%%%%%%%%%%%
%%%%%%%%%%%%%%%%%%%%%%%%%%%%%%%%%%%%%%%%%%%%%%%%%%%%%%%%%%%%%%%%%%%%%%%%%%%%%%%
%%%%%%%%%%%%%%%%%%%%%%%%%%%%%%%%%%%%%%%%%%%%%%%%%%%%%%%%%%%%%%%%%%%%%%%%%%%%%%%
%%%%%%%%%%%%%%%%%%%%%%%%%%%%%%%%%%%%%%%%%%%%%%%%%%%%%%%%%%%%%%%%%%%%%%%%%%%%%%%

It is also conceivable for a given repulsive interaction to support excitonic and superconducting states, described by the fermion bilinears $\sum^{K}_{i=1} \Psi^\dag_{\rm Nam} R^{\rm exc}_i \Psi_{\rm Nam}$ and $\sum^{L}_{i=1} \Psi^\dag_{\rm Nam} R^{\rm pair}_i \Psi_{\rm Nam}$, respectively. These two orders respectively transform as vectors under O$(K)$ and O$(L)$ rotations, constituted by $K$ and $L$ number of mutually anticommuting matrices. Then the pairing and excitonic states are realized at low and high temperatures, respectively, for the same repulsive four-fermion interaction if the corresponding vector order parameters ($\vec{R}^{\rm exc}$ and $\vec{R}^{\rm pair}$) form composite vectors under the O$(N)$ rotations, where $K,L < N \leq K+L$ (SR3). For all the interaction channels that support both excitonic and pairing states, SR3 remains operative therein. See Table~\ref{tab:individual_channels}.

We also notice that when SR3 is operative, the pairing state emerges at low temperatures, while the excitonic order appears in the phase diagram at a higher temperature. Otherwise, these two orders are respectively realized for weaker and stronger repulsive interactions for the following reason. For weak repulsive interactions, the existence of the Fermi surface tames its natural tendency to favor excitonic order, rather boosts the propensity toward the appropriate pairing order, following SR2. Eventually for stronger repulsion, the propensity toward the pairing order due to the Fermi surface becomes insufficient, and the system then prefers the formation of an excitonic order, following SR1 or SR2. Furthermore, a Fermi surface is maximally gapped by superconductors, yielding optimal gain of the condensation energy. As at low temperature the condensation energy gain dominates over the entropy in the free energy, the nucleation of a superconducting ground state is then favored. On the other hand, an excitonic order in the presence of an underlying Fermi surface supports gapless fermionic excitations yielding a larger entropy. Consequently, excitonic or particle-hole orders appear at higher temperature. These outcomes follow the general (qualitative) arguments of the organizing principle. It should, however, be noted that repulsive electron-electron interactions in some specific channels do not support any paired state and only favors particle-hole orders, which still follow SR1 or SR2, see Table~\ref{tab:individual_channels}.

We exemplify these SRs and the organizing principle with repulsive interaction in a specific channel $g_{_\perp}^s$ with $g_{_\perp}^s>0$. As a reference consult the second row of Table~\ref{tab:individual_channels}. For such a choice of the interaction channel $\vec{Q}=(\Gamma_{301},\Gamma_{302})$, and the superconducting order realized at low temperature is the $s$-wave pairing ($\Delta^1_p$) with $\vec{R}=(\Gamma_{100},\Gamma_{200})$. In this case, SR2 becomes operative with $A_M=2$ for each entry of $\vec{R}$. The excitonic order realized at higher temperature is the spin current ($\Delta_{_\perp}^s$) with $\vec{R}=(\Gamma_{301},\Gamma_{302})$ following SR1 ($\vec{R} \equiv \vec{Q}$). In this case, $K=2$ and $L=2$, and the composite order transforms as a vector under O($N$) rotation with $N=4$, thus following SR3 as $K,L < N = K+L$ in this case. Explicitly, the composite O(4) vector order parameter is given by $(\Gamma_{301},\Gamma_{302}, \Gamma_{100},\Gamma_{200})$. In the $(g_{_\perp}^s, t)$ plane, while the superconducting order nucleates at lower temperature and weaker coupling, the excitonic state sets in at a higher temperature and stronger coupling, as shown in Figs.~\ref{fig:QSHIPD1}(d)-\ref{fig:QSHIPD1}(f), following the organizing principle. Readers are encouraged to check that these set of rules remain operative for repulsive interactions in any channel.

\section{Hubbard model}~\label{sec:Hubbard}

%%%%%%%%%%%%%%%%%%%%%%%%%%%%%%%%%%%%%%%%%%%%%%%%%%%%%%%%%%%%%%%%%%%
%%%%%%%%%%%%%%%%%%%%%%%%%%%%%%%%%%%%%%%%%%%%%%%%%%%%%%%%%%%%%%%%%%%
%%%%%%%%%%%%%%%%%%%%%%%%%%%%%%%%%%%%%%%%%%%%%%%%%%%%%%%%%%%%%%%%%%%
%%%%%%%%%%%%%%%%%%%%%%%%%%%%%%%%%%%%%%%%%%%%%%%%%%%%%%%%%%%%%%%%%%%
%%%%%%%%%%%%%%%%%%%%%%%%%%%%%%%%%%%%%%%%%%%%%%%%%%%%%%%%%%%%%%%%%%%
\begin{figure*}[t!]
\includegraphics[width=1.00\linewidth]{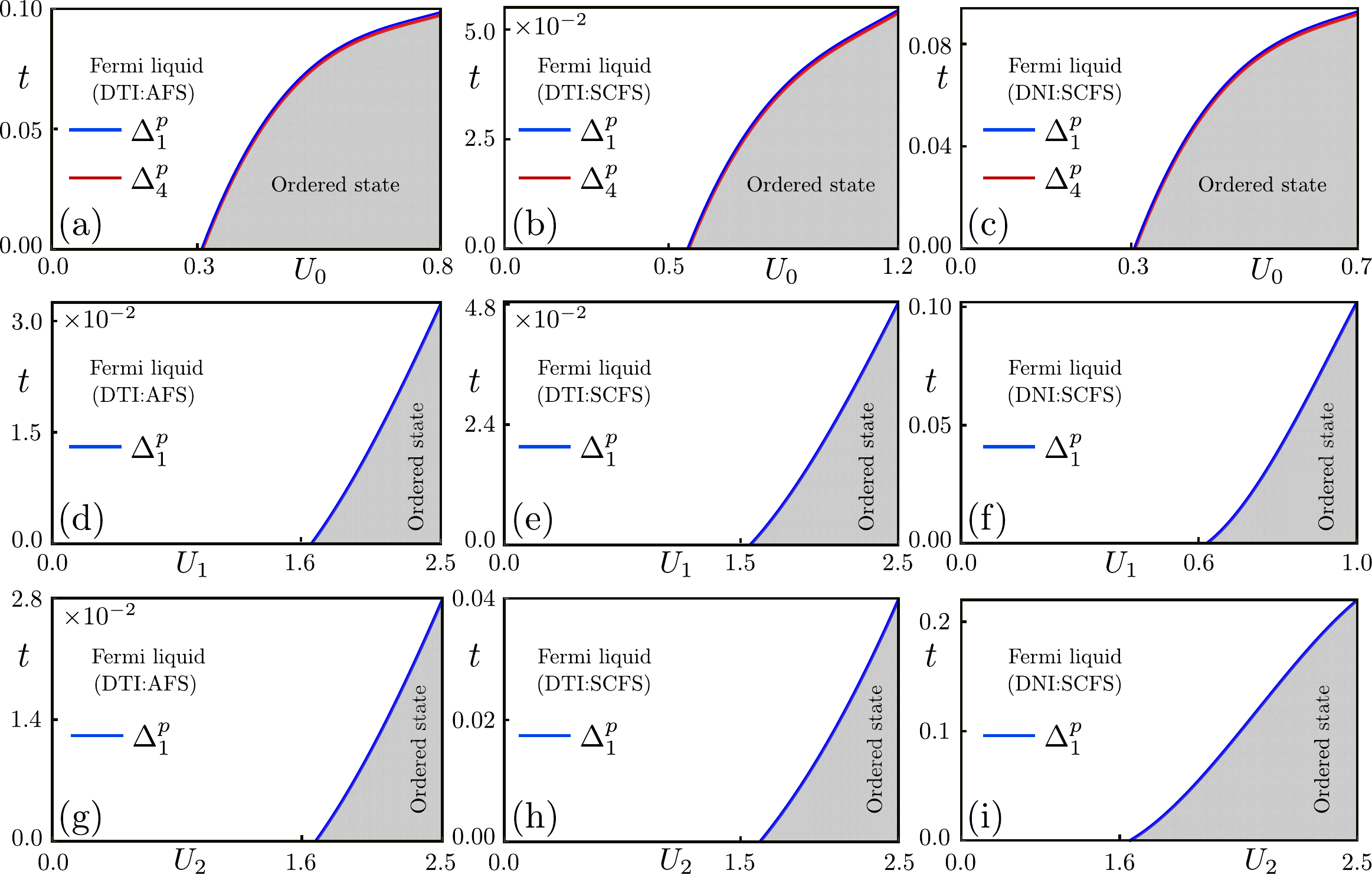}
\caption{The phase diagrams for [(a)-(c)] orbital Hubbard repulsion with strength $U_0$, [(d)-(f)] spin Hubbard repulsion with spin projection along the $z$ direction with strength $U_1$, and [(g)-(i)] spin Hubbard repulsion with spin projection confined within the easy or $xy$ plane with strength $U_2$ in a time-reversal symmetric doped planar Dirac insulator from class AII for $\alpha=0.5$. See Sec.~\ref{sec:Hubbard} for details. The parameter values $(m,b,\mu)$ are equal to
(a) $(-0.9,1.0,0.86)$, (b) $(-0.9,1.0,0.91)$, (c) $(0.05,0.05,0.86)$, 
(d) $(-0.7,3.8,0.55)$, (e) $(-0.4,3.8,0.45)$, (f) $(0.1,1.5,0.25)$, 
(g) $(-0.7,3.8,0.55)$, (h) $(-0.4,3.8,0.45)$, and (i) $(0.1,1.5,0.25)$. 
While orbital Hubbard repulsion causes a degenerate nucleation of even-parity and odd-parity paired states, denoted by $\Delta^p_1$ and $\Delta^p_4$, respectively, the spin Hubbard repulsions prefer the former paired state. Around the Fermi surface $\Delta^p_4$ ($\Delta^p_1$) pairing assumes the form of a topological $p \pm ip$ (conventional $s$-wave) pairing. See Appendix~\ref{append:bandprojection} for details. 
The rest of the details are same as in Figs.~\ref{Fig:QAHIPD} and~\ref{fig:QSHIPD1}. Dimensionless coupling constants ($U_0$, $U_1$, and $U_2$) are measured in units of $\epsilon$, where $\epsilon=d-1$.
}~\label{fig:QSHIPDHubbard}
\end{figure*}
%%%%%%%%%%%%%%%%%%%%%%%%%%%%%%%%%%%%%%%%%%%%%%%%%%%%%%%%%%%%%%%%%%%
%%%%%%%%%%%%%%%%%%%%%%%%%%%%%%%%%%%%%%%%%%%%%%%%%%%%%%%%%%%%%%%%%%%
%%%%%%%%%%%%%%%%%%%%%%%%%%%%%%%%%%%%%%%%%%%%%%%%%%%%%%%%%%%%%%%%%%%
%%%%%%%%%%%%%%%%%%%%%%%%%%%%%%%%%%%%%%%%%%%%%%%%%%%%%%%%%%%%%%%%%%%
%%%%%%%%%%%%%%%%%%%%%%%%%%%%%%%%%%%%%%%%%%%%%%%%%%%%%%%%%%%%%%%%%%%

We now substantiate the emergent superconductivity in class A and class AII interacting doped Dirac insulators within the framework of appropriate repulsive Hubbard models. The parity or orbital degrees of freedom allows us to consider on-site repulsion of strength $U_0 (>0)$ among spinless or spin-polarized fermions with opposite-parity eigenvalues in a class A system. The corresponding Hamiltonian reads 
\begin{equation}
H^{\rm orb, A}_{\rm Hubb} =\frac{U_0}{2} \sum_{i} n_{+,i} \;  n_{-,i},
\end{equation}
where $n_{\tau,i}$ is the fermionic density at site $i$ with parity $\tau=\pm$. In the spinor basis from Sec.~\ref{sec:dopedQAHI}, the above Hamiltonian can be written as 
\begin{equation}
H^{\rm orb, A}_{\rm Hubb}=\frac{U_0}{8} \left[ \left( \psi^\dagger \tau_0 \psi\right)^2 - \left( \psi^\dagger \tau_3 \psi\right)^2 \right]
\equiv \frac{U_0}{4} \; \left( \psi^\dagger \tau_0 \psi\right)^2.
\end{equation}
While arriving at the final expression we made use of the Fierz relation. Hence, the phase diagrams of the orbital Hubbard model in doped class A Dirac insulator are obtained with the initial condition $g_{_0}(0)=U_0/4$ and are thus identical to the ones displayed in Fig.~\ref{Fig:QAHIPD}, once we identify the horizontal axis as $U_0/4$. Therefore, on-site Hubbard repulsion among spinless or spin polarized fermions supports the topological pairing ($\Delta_p$) (see Appendix~\ref{append:bandprojection}) in a two-dimensional doped Dirac insulator of class A irrespective of the Fermi surface geometry.

The orbital Hubbard Hamiltonian in class AII systems containing both orbital and spin degrees of freedom takes the following form 
\begin{equation}
H^{\rm orb, AII}_{\rm Hubb} =\frac{U_0}{2} \sum_{i} \sum_{\sigma=\uparrow, \downarrow} n_{+,\sigma,i} \;  n_{-,\sigma,i},
\end{equation} 
where $n_{\tau,\sigma,i}$ is the fermionic density at site $i$ with spin projection $\sigma$, and with parity $\tau$. In the announced spinor basis from Sec~\ref{sec:dopedQSHI}, we find 
\begin{equation}
H^{\rm orb, AII}_{\rm Hubb} =\frac{U_0}{4} \left[ \left( \psi^\dagger \Gamma_{00} \psi\right)^2 - \left( \psi^\dagger \Gamma_{03} \psi \right)^2 \right].
\end{equation} 
Hence, the phase diagram of the orbital Hubbard model in this system can be obtained with the initial conditions $g^s_{_0}(0)=U_0/4$ and $g^s_{_3}(0)=-U_0/4$. Notice that the repulsive (attractive) interaction in the $g^s_{_0}$ ($g^s_{_3}$) favors the nucleation of even- (odd-) parity $\Delta^p_1$ ($\Delta^p_4$) pairing. Since the strengths of these two interaction channels have equal magnitude in $H^{\rm orb, AII}_{\rm Hubb}$, the orbital Hubbard repulsion ($U_0>0$) fosters a degenerate condensation of these two paired states irrespective of the geometry of the underlying Fermi surface, as shown in Figs.~\ref{fig:QSHIPDHubbard}(a)-(c).

Finally, we turn our focus on the conventional Hubbard interaction operative among fermions living on the same lattice site but possessing opposite spin projections. The corresponding Hamiltonian is given by 
\begin{equation}
H^{\rm spin, AII}_{\rm Hubb}= \frac{U}{2} \sum_{i} \sum_{\tau=+, -} n_{\tau,\uparrow,i} \;  n_{\tau,\downarrow,i},
\end{equation}
where $U(>0)$ denotes the strength of onsite Hubbard repulsion. If we choose the spin quantization axis along the $z$ axis, then the above model in the spinor basis, introduced in Sec.~\ref{sec:dopedQSHI}, reads as 
\begin{equation}~\label{eq:HubbardZ}
H^{\rm spin, AII}_{\rm Hubb}=\frac{U}{4} \left[ \left( \psi^\dagger \Gamma_{00} \psi\right)^2 - \left( \psi^\dagger \Gamma_{30} \psi \right)^2 \right].
\end{equation} 
However, this representation does not manifest the SU(2) spin rotational symmetry, which can be recovered by using Fierz relations involving only the spin indices [analogously to Eq.~\eqref{eq:ftzconstraintsQAHI} after taking $\tau_\alpha \to \sigma_\alpha$ for $\alpha=0,1,2,3$ therein], leading to 
\begin{equation}~\label{eq:HubbardSU2}
H^{\rm spin, AII}_{\rm Hubb}=\frac{U}{8} \left[ \left( \psi^\dagger \Gamma_{00} \psi\right)^2 - \sum_{j=1,2,3} \left( \psi^\dagger \Gamma_{j0} \psi \right)^2 \right].
\end{equation} 
As class AII does not possess the spin SU(2) symmetry, we need to split the second term into the easy-axis ($j=3$) and easy-plane ($j=1,2$) components. This splitting leaves us with an ambiguity on how to distribute the density-density interaction $\left( \psi^\dagger \Gamma_{00} \psi\right)^2$ between these two sectors, leading to the following proposed ansatz for $H^{\rm spin, AII}_{\rm Hubb}$ in class AII system
\begin{eqnarray}~\label{eq:HubbardAII}
H^{\rm spin, AII}_{\rm Hubb} &=& \frac{U_1}{8} \left[ (1-\alpha) \left( \psi^\dagger \Gamma_{00} \psi\right)^2 - \left( \psi^\dagger \Gamma_{30} \psi \right)^2 \right] \nonumber \\
&+& \frac{U_2}{8} \left[ \alpha \left( \psi^\dagger \Gamma_{00} \psi\right)^2 - \sum_{j=1,2} \left( \psi^\dagger \Gamma_{j0} \psi \right)^2 \right].\:\:\:
\end{eqnarray}
Here $\alpha$ is an arbitrary real parameter with $0 \leq \alpha \leq 1$, and for $U_1=U_2=U$ we recover the SU(2) symmetric form of the Hubbard model for any $\alpha$, shown in Eq.~\eqref{eq:HubbardSU2}. Notice that upon splitting the SU(2) symmetric Hubbard model into the easy-axis and easy-plane components we assign them two different independent bare coupling constants $U_1$ and $U_2$, respectively. Even if we set $U_1(0)=U_2(0)$, under the coarse graining these two sectors acquire distinct coupling constant strengths. In terms of four linearly independent coupling constants, see Sec.~\ref{sec:dopedQSHI}, the above Hubbard model takes the form   
\begin{eqnarray}
&&H^{\rm spin, AII}_{\rm Hubb} = \frac{U_1}{8} \left[ (2-\alpha) (\psi^{\dagger}\Gamma_{00}\psi)^2+ \sum_{j=0,3} (\psi^{\dagger}\Gamma_{j3}\psi)^2 \right] \nonumber \\
&+&  \frac{U_2}{8} \bigg[ (1+\alpha) (\psi^{\dagger}\Gamma_{00}\psi)^2
+ \sum_{j=1,2} (\psi^{\dagger}\Gamma_{0j}\psi)^2 
- (\psi^{\dagger}\Gamma_{33}\psi)^2 \bigg]. \nonumber \\
\end{eqnarray}

Therefore, the phase diagrams of easy-axis Hubbard model can be obtained with the initial conditions
\begin{equation}
g^s_{_0}(0)=(2-\alpha) \frac{U_1}{8}, \; 
g^s_{_3}(0)= \frac{U_1}{8}, 
\;\; \text{and} \;\;
g^z_{_3}(0)= \frac{U_1}{8}. 
\end{equation} 
Notice the repulsive interaction in the $g^s_{_0}$ and $g^z_{_3}$ channels favor $\Delta^p_1$ pairing, and repulsive $g^s_{_3}$ interaction favors $\Delta^p_2$ pairing. With the above initial conditions the propensity toward the nucleation of the $\Delta^p_1$ paired state is always stronger than that in the $\Delta^p_2$ channel. Consequently, repulsive easy-axis Hubbard interaction ($U_1>0$) is conducive to the condensation of even-parity $\Delta^p_1$ paired state for any $\alpha$ within the range $0 \leq \alpha \leq 1$ irrespective of the Fermi surface topology, as shown in Figs.~\ref{fig:QSHIPDHubbard}(d)-(f).

By contrast, the phase diagrams for the easy-plane Hubbard model is obtained with the initial conditions 
\begin{equation}
g^s_{_0}(0)=(1+\alpha) \frac{U_2}{8}, \;
g^s_{_\perp}(0)= \frac{U_2}{8}, 
\;\; \text{and} \;\;
g^z_{_3}(0)= -\frac{U_2}{8}.
\end{equation}
We note that repulsive interaction in the $g^s_{_0}$ and $g^s_{_\perp}$ channels favor $\Delta^p_1$ pairing, while an attractive interaction in the $g^z_{_3}$ channel supports $\Delta^p_4$ pairing. With the above initial condition and for any $\alpha$ within the range $0 \leq \alpha \leq 1$, the propensity toward the nucleation of the $\Delta^p_1$ pairing dominates. Hence, easy-plane repulsive Hubbard interaction ($U_2>0$) also favors the $\Delta^p_1$ paired state for any Fermi surface geometry, as shown in Fig.~\ref{fig:QSHIPDHubbard}(g)-(i).

\section{Summary and discussions}~\label{sec:summary}

To summarize, we show that purely repulsive electron-electron interactions in two-dimensional doped Dirac insulators (topological or normal), belonging to class A and class AII, can favor condensation of Cooper pairs in various symmetry allowed superconducting channels which include the topological ones as well. The nature of the paired state is insensitive to the geometry of the underlying Fermi surface (annular or simply connected) and is tied to the symmetry of the dominant repulsive interaction following a certain set of selection rules. Furthermore, we find that on-site Hubbard repulsion ($U_0>0$) among fermions with opposite-parity eigenvalues living on the same site of the lattice is conducive to the formation of topological superconductors. Specifically, in doped class A and class AII Dirac insulators they take the form of $p+ip$ and $p \pm ip$ pairings, respectively, belonging to the class D and class DIII. We arrive at these conclusions by (a) performing an unbiased RG analysis to the leading order that is controlled by a small parameter $\epsilon$, measuring the distance from the lower critical one spatial dimension where the four-fermion interactions become marginal with $\epsilon=d-1$, and (b) projecting all the symmetry allowed local paired states onto the Fermi surface, realized within the conduction band, for example.

Remarkably doped class AII Dirac insulators also accommodate topological pairing belonging to class C for a repulsive interaction in a specific channel, namely $g^s_{_3}$. Unfortunately, we have not found the trace of such a paired state within the framework of the repulsive Hubbard model. Nevertheless, our formalism is sufficiently general to account for any microscopic finite-range interactions that can be captured by appropriate initial conditions for four linearly independent local quartic terms. In the future it will be worth searching for a microscopic interacting model that can in principle foster class C topological pairing in this system. Notice that in two spatial dimensions, the ten-fold classification allows five nontrivial topological classes. Two of them correspond to insulators (class A and class AII) and the remaining three to superconductors (class D, class DIII, and class C). Fascinatingly, the symmetry allowed local pairings in these two classes of Dirac insulators harbor topological pairing from all three superconducting classes.

A comment regarding the stability of various ordered states at finite temperature is due at this stage, due to the reduced dimensionality of the systems we are considering here. Notice that only discrete (such as Ising or $Z_2$) symmetry breaking states are stable at finite $T$ and develop long-range order in two dimensions through a true finite temperature phase transition. Despite being accompanied by one Nambu-Goldstone mode, O(2) symmetry breaking orders also show finite-temperature phase transition, which is, however, Kosterliz-Thouless in nature and the ordered states display quasi-long-range algebraic order~\cite{KT:original}, as is the case for the superconducting orders, for example. Any order that breaks a O($n$) symmetry with $n \geq 3$ does not show true long-range order or true finite-temperature phase transition (Mermin-Wagner theorem)~\cite{goldstone:1, goldstone:2, MarminWagner:1, MarminWagner:2, MarminWagner:3}. In two systems, we study here (class A and class AII), there is no such order, however (see Tables~\ref{tab:ClassA} and~\ref{tab:bilinears}). It should, nonetheless, be noted that one-loop RG analysis performed here about the disordered Fermi liquid state cannot capture these well-known results from the Mermin-Wangner theorem. Construction of a nonlinear sigma model within the ordered phase, obtained by integrating out gapped fermionic degrees of freedom, immediately leads to these conclusions (except the Kosterlitz-Thouless transition~\cite{KT:original}).

Our results can be experimentally relevant in doped quantum anomalous Hall insulators, realizable in thin films of magnetically doped (by Cr or V or Fe, for example) three-dimensional topological insulators, such as Bi$_2$Se$_3$, Bi$_2$Te$_3$, and Sb$_2$Te$_3$~\cite{QAHI:1, QAHI:2, QAHI:3}, and quantum spin Hall insulators in CdTe-HgTe~\cite{TITSC:8, QSHI:2} and InAs-SbTe~\cite{QSHI:3} quantum wells. Also, a jacutingaite compound Pt$_2$HgSe$_3$ has been identified as a quantum spin Hall insulator~\cite{QSHI:4, QSHI:5, QSHI:6}, which can foster superconductivity at low temperature~\cite{QSHI:7}. The two-dimensional nature of these materials should permit a controlled chemical doping of desired carrier density by tuning the gate voltage without compromising with the sample quality (i.e., without introducing random charge impurities) to foster Fermi surfaces therein, necessary for superconductivity to set in. The predicted possibility of harnessing topological superconductors on such material platforms should therefore stimulate future experiments in this direction. Our work should also trigger a search for quantum anomalous and spin Hall insulators in strongly correlated materials (such as the Kondo systems). When doped, these systems will possibly constitute a promising landscape for topological superconductivity due to the strong Hubbardlike interactions therein. Our proposal to search for planar odd-parity topological superconductors in doped quantum anomalous and spin Hall insulator materials can be justified phenomenologically from the observed superconductivity in three-dimensional intercalated (Cu$_x$Bi$_2$Se$_3$) and
doped (Sn$_{1-x}$In$_x$Te) topological insulators below a few Kelvin. Such paired states feature a surface zero-bias-conductance peak, a possible hallmark of the gapless surface Majorana fermions, suggesting topological nature of the underlying odd-parity paired state~\cite{3DTSC:1, 3DTSC:2, 3DTSC:3, 3DTSC:4}.

Promisingly, the Qi-Wu-Zhang model~\cite{QWZ, QAHIOL:1} and Haldane model~\cite{haldane, QAHIOL:2}, both representing lattice realizations of quantum anomalous Hall insulators, have recently been realized on optical lattices. Notice that the massive Dirac Hamiltonian in class A, employed in this work, is obtained from the lattice-regularized Qi-Wu-Zhang model in the continuum limit. On optical lattices the doping level and the strength of Hubbardlike local or short-range interactions can be tuned efficiently. On this platform the Hubbard model has been simulated on two-dimensional square lattice to map its global phase diagram, featuring antiferromagnetism at and close to the half-filling and superconductivity away from it~\cite{OpLat:1, OpLat:2, OpLat:3, OpLat:4}, mimicking the cartoon phase diagram from Fig.~\ref{fig:philosophy}(a). Nowadays three-dimensional optical cubic lattice can be realized as well~\cite{OpLat:5}. Therefore, optical lattices of spin-orbit coupled ultracold fermions also constitute a promising platform where our proposed phase diagram from Fig.~\ref{fig:philosophy}(b) fostering topological or normal Dirac insulators and emergent topological superfluids of neutral fermions can be observed experimentally.

Although the transition temperature for superconductivity ($t_c$) can be crudely estimated from our RG calculations, readers must realize that such estimations can easily be off by several factors (see, for example Ref.~\cite{KL:1}) as they depend on many nonuniversal (often not known accurately) parameters of the system. We measure the dimensionless coupling constants ($g$) and the transition temperature ($t_c$) in units of $E_\Lambda=\Lambda v$ that sets an energy scale over which the band dispersion can be described by massive Dirac fermions approximately. Typically, in topological semiconductors $E_\Lambda \sim 1$ eV. Then from the phase diagram in Fig.~\ref{Fig:QAHIPD}(a), we find $t_c=500$ mK for the \emph{bare} Hubbard interaction $1.2$ eV, which is not too unreasonable for this pairing to be potentially observed in some of the candidate real materials.

The emergent topological $p+ip$ ($p \pm ip$) paired state can be identified from half-quantized thermal Hall conductivity $\kappa_{xy}$ (quantized longitudinal thermal conductance $G^{th}_{xx}$) in units of $\kappa_0=\pi^2 k^2_B T/(3h)$, where $k_B$ ($h$) is the Boltzmann (Planck) constant~\cite{TSCRes:1, TSCRes:2, TSCRes:3, TSCRes:4}. By contrast, the class C paired state manifests a quantized $\kappa_{xy}$~\cite{TSCRes:4, TSCRes:5, TSCRes:6}. Recent time has witnessed tremendous experimental progress leading to successful measurements of (half)quantized thermal responses in sufficiently clean quantum materials~\cite{Thermal:1, Thermal:2, Thermal:3, Thermal:4, Thermal:5}. On optical lattices, quantized transports can be measured from the ``heating effect", for example~\cite{Thermal:6}. Finally, scanning tunneling microscope can be instrumental to identify emergent topological superconductors from their signature Majorana edge modes in quantum crystals~\cite{STM:1, STM:2}. Analogously the Majorana edge modes on optical lattices can be probed from their local density states by local radio frequency spectroscopy, for example~\cite{LDOSOC:1, LDOSOC:2, LDOSOC:3, LDOSOC:4}. Therefore, identification of planar topological superconductors in two-dimensional correlated doped Dirac insulators should be within the reach of currently available experimental facilities and techniques.

\acknowledgments

This work was supported by Dr.\ Hyo Sang Lee Graduate Fellowship from Lehigh University (S.A.M.), NSF CAREER Grant No.\ DMR-2238679 (S.A.M.\ and B.R.), and the Startup Grant of B.R.\ from Lehigh University (S.K.D). S.A.M.\ thanks Andr\'{a}s L. Szab\'{o} for valuable discussions. We are thankful to Vladimir Juri\v{c}i\'c for insightful discussions. We also acknowledge Christopher A.\ Leong for valuable inputs on the manuscript.

\appendix

\section{Fierz reductions}~\label{append:fierz}

In this Appendix, we show the Fierz reduction of the number of linearly independent local quartic terms. The Fierz identity allows us to write any four-fermion term as the linear combination of the others expressed in terms of $2^{2N}$ number of generators of the U($2^N$) group as follows 
\allowdisplaybreaks[4]
\begin{eqnarray}~\label{SMeq:ftzmat}
    &&[\psi^\dagger (x) M \psi(x)] [\psi^\dagger (y) N \psi(y)]= -\frac{1}{2^{2N}} \sum_{a,b} {\rm Tr}[M \Gamma^b N \Gamma^a] \nonumber \\
		&\times& [\psi^\dagger(x) \Gamma^a \psi (y)] [\psi^\dagger(y) \Gamma^b \psi (x)].
\end{eqnarray}
Here $M$ and $N$ are $2^N$-dimensional Hermitian matrices, and $\Gamma^a$ are the generators of $U(2^N)$ group. Here we apply this identity for contact local interactions with $x=y$.

In a time-reversal symmetry breaking insulator, considered in Sec.~\ref{sec:dopedQAHI}, there are altogether three local quartic interaction terms that can be organized in terms of the elements of a vector $X$, given by 
\begin{equation}~\label{eq:ftzbasisQAHI}
        X^\top = \bigg( (\psi^\dagger \tau_{0}\psi)^2, 
				\sum_{j=1,2} (\psi^\dagger \tau_{j}\psi)^2,
				(\psi^\dagger \tau_{3}\psi)^2 \bigg),
\end{equation}
where $\top$ stands for transposition. Then the above Fierz constraint with $N=1$ can be cast as a matrix equation $F X=0$, where the Fierz matrix $F$ is given by 
\begin{equation}
F=\left( \begin{array}{ccc}
3 & 2 & 1  \\
1 & 2 & -1 \\
1 &-2 & 3 \\ 
\end{array} \right).
\end{equation}
The number of linearly independent quartic terms is equal to the difference between the dimensionality of $F$, denoted by ${\rm dim}(F)$ and its rank, denoted by ${\rm rank}(F)$. For the above Fierz matrix ${\rm dim}(F)=3$ and ${\rm rank}(F)=2$. Hence, the number of linearly independent quartic term is one, which we choose to be $(\psi^\dagger \tau_{0}\psi)^2$ without any loss of generality. The remaining two quartic terms are then given by 
\allowdisplaybreaks[4]
\begin{eqnarray}~\label{eq:ftzconstraintsQAHI}
\sum_{j=1,2} (\psi^\dagger\tau_{j}\psi)^2 = -  (\psi^\dagger\tau_{0}\psi)^2
\: \text{and} \:
 (\psi^\dagger\tau_{3}\psi)^2 = -(\psi^\dagger \tau_{0}\psi)^2. \nonumber \\
\end{eqnarray}
Therefore, whenever these two quartic terms are generated during coarse graining they can be expressed in terms of $(\psi^\dagger \tau_{0}\psi)^2$. So, the interacting Lagrangian in terms of this single quartic term remains closed under RG to any order in the perturbation theory.

We now proceed with a similar approach to find the linearly independent quartic terms in a system describing a time-reversal symmetric insulator, considered in Sec.~\ref{sec:dopedQSHI}. Now nine quartic terms are organized as 
\begin{eqnarray}
&&X^\top = \bigg( (\psi^{\dagger} \Gamma_{00}\psi)^2 , \sum_{j=1,2} (\psi^{\dagger} \Gamma_{0j}\psi)^2, 
(\psi^{\dagger} \Gamma_{03}\psi)^2, \nonumber \\ 
&& (\psi^{\dagger} \Gamma_{30}\psi)^2, \sum_{j=1,2} (\psi^{\dagger} \Gamma_{3j}\psi)^2, (\psi^{\dagger} \Gamma_{33}\psi)^2, 
\sum_{j=1,2}(\psi^{\dagger} \Gamma_{j0}\psi)^2, \nonumber \\
&& \sum_{j=1,2} [(\psi^{\dagger} \Gamma_{j1}\psi)^2+(\psi^{\dagger} \Gamma_{j2}\psi)^2],
\sum_{j=1,2} (\psi^{\dagger} \Gamma_{j3}\psi)^2
\bigg). 
\end{eqnarray}
The corresponding Fierz matrix with ${\rm dim}(F)=9$ and $N=2$ reads as 
\begin{equation}
F=\left(
\begin{array}{ccccccccc}
 5 & 1 & 1 & 1 & 1 & 1 & 1 & 1 & 1 \\
 2 & 4 & -2 & 2 & 0 & -2 & 2 & 0 & -2 \\
 1 & -1 & 5 & 1 & -1 & 1 & 1 & -1 & 1 \\
 1 & 1 & 1 & 5 & 1 & 1 & -1 & -1 & -1 \\
 2 & 0 & -2 & 2 & 4 & -2 & -2 & 0 & 2 \\
 1 & -1 & 1 & 1 & -1 & 5 & -1 & 1 & -1 \\
 2 & 2 & 2 & -2 & -2 & -2 & 4 & 0 & 0 \\
 4 & 0 & -4 & -4 & 0 & 4 & 0 & 4 & 0 \\
 2 & -2 & 2 & -2 & 2 & -2 & 0 & 0 & 4 \\
\end{array}
\right).
\end{equation}  
The rank of this matrix is five [${\rm rank}(F)=5$] and therefore the number of linearly independent quartic terms is four, which we choose to be 
\begin{equation}
(\psi^{\dagger} \Gamma_{00}\psi)^2,\; 
\sum_{j=1,2}(\psi^{\dagger} \Gamma_{0j}\psi)^2,\;
(\psi^{\dagger} \Gamma_{03}\psi)^2, 
\: \text{and} \:
(\psi^{\dagger} \Gamma_{33}\psi)^2. \nonumber 
\end{equation} 
Hence, the interacting Lagrangian in terms of these four quartic terms remains closed to any order in the perturbation theory as whenever any one of the remaining five quartic terms is generated during coarse graining it can be written as a linear combination of the chosen ones. Specifically, the linear relationships between these four quartic terms and the remaining ones can be cast as 
\begin{eqnarray} 
&& \begin{pmatrix}
(\psi^{\dagger} \Gamma_{30}\psi)^2\\
(\psi^{\dagger} \Gamma_{31}\psi)^2+(\psi^{\dagger} \Gamma_{32}\psi)^2 \\
(\psi^{\dagger} \Gamma_{10}\psi)^2 + (\psi^{\dagger} \Gamma_{20}\psi)^2 \\
(\psi^{\dagger} \Gamma_{11}\psi)^2+(\psi^{\dagger} \Gamma_{12}\psi)^2 
+ (\psi^{\dagger} \Gamma_{21}\psi)^2+(\psi^{\dagger} \Gamma_{22}\psi)^2\\
(\psi^{\dagger} \Gamma_{13}\psi)^2 + (\psi^{\dagger} \Gamma_{23}\psi)^2 \\
\end{pmatrix} \nonumber \\
&=& \begin{pmatrix}
-1&0&-1&-1\\
0&-1&2&2\\
-1&-1&0&1\\
-2&0&0&-2\\
-1&1&-2&-1\\
\end{pmatrix}
\begin{pmatrix}
(\psi^{\dagger} \Gamma_{00}\psi)^2\\
(\psi^{\dagger} \Gamma_{01}\psi)^2+(\psi^{\dagger} \Gamma_{02}\psi)^2 \\
(\psi^{\dagger} \Gamma_{03}\psi)^2\\
(\psi^{\dagger} \Gamma_{33}\psi)^2
\end{pmatrix}.
\end{eqnarray}

\section{Band projection and emergent topology}~\label{append:bandprojection}

In this Appendix, we identify the emergent topology of local pairings by projecting them near the Fermi surface realized on the conduction band (for $\mu>0$). We begin this discussion by focusing on class A doped Dirac insulators. We choose to work in a slightly different Nambu-doubled basis, defined as $\psi^\top_{\rm Nam}=(\psi_{\omega_n,k},\psi_{-\omega_n,-k}^\star)$ for technical ease, than the one introduced in Sec.~\ref{sec:dopedQAHI}. In this basis the free-fermion massive Dirac Hamiltonian and the one describing the paired state read as 
\begin{eqnarray}
\hat{h}^{\rm Nam}_{\rm Dir, A} &=& v \left( \Gamma_{01} k_x + \Gamma_{32} k_y \right) + (m+b k^2) \Gamma_{33} -\mu \Gamma_{30} \nonumber \\
\text{and} \:\: \hat{h}^{\rm A}_{\rm pair} &=& \Delta_p \left[ \cos(\phi_{\rm sc}) \Gamma_{12} + \sin(\phi_{\rm sc}) \Gamma_{22} \right], 
\end{eqnarray}
respectively, obtained from Eqs.~\eqref{HDirNamA} and~\eqref{eq:hampairQAHI} after performing a unitary rotation by $\tau_0 \oplus \tau_1$, where $\phi_{\rm sc}$ is the U(1) superconducting phase. The unitary matrix that diagonalizes $\hat{h}^{\rm Nam}_{\rm Dir}$ is obtained by columnwise arranging its eigenvectors, given by $U^{\rm A}_{\rm Nam}=U(k_x,k_y,m_{\vec{k}}) \oplus U(k_x,-k_y,-m_{\vec{k}})$, where  
\begin{eqnarray}
&&U (k_x, k_y, m) = e^{-i\phi_{\vec{k}}} \\
&\times& \begin{pmatrix}
                    \frac{E_k+(m+bk^2)}{\sqrt{2 E_k(E_k+(m+bk^2))}} &-\frac{E_k-(m+bk^2)}{\sqrt{2E_k(E_k-(m+bk^2))}}\\
                  \frac{v(k_x+ik_y)}{\sqrt{2E_k(E_k+(m+bk^2))}} & \frac{v(k_x+ik_y)}{\sqrt{2E_k(E_k-(m+bk^2))}}\\
                \end{pmatrix}, \nonumber 
\end{eqnarray}
$\phi_{\vec{k}}=\tan^{-1}(k_y/k_x)$, $m_{\vec{k}}=m+bk^2$, and $E_k=[v^2 k^2 + (m+b k^2)^2]^{1/2}$. We then find 
\begin{equation}
\left( U^{\rm A}_{\rm Nam} \right)^\dagger \hat{h}^{\rm Nam}_{\rm Dir, A} U^{\rm A}_{\rm Nam}={\rm diag}.(E_k,-E_k,E_k,-E_k)-\mu \Gamma_{30}.
\end{equation}
Hence, the reduced Hamiltonian describing a collection of gapless excitations around the Fermi surface within the conduction band is given by 
\begin{equation}
\hat{h}^{\rm cond}_{\rm FS, A}= \left( E_k-\mu \right)\eta_3 \approx \left( \frac{v^2 k^2}{2 m} -\tilde{\mu} \right) \eta_3 + {\mathcal O} (k^4).
\end{equation}
While arriving at the last expression we performed a large mass expansion of $E_k$, and introduced a rescaled chemical potential $\tilde{\mu}=\mu-m$, measured from the bottom of the conduction band. The associated Nambu-doubled basis read as $\Psi^\top_{\rm Nam}=(c_{\vec{k},\rm CB}, c^\dagger_{-\vec{k},\rm CB})$, where $c_{\vec{k},\rm CB}$ ($c^\dagger_{\vec{k},\rm CB}$) is the fermionic annihilation (creation) operator on the conduction band (CB) with momentum $\vec{k}$.

The diagonalization procedure when applied on the pairing Hamiltonian $\hat{h}_{\rm pair}$ yields 
\begin{eqnarray}
&&\left( U^{\rm A}_{\rm Nam} \right)^\dagger \hat{h}^{\rm A}_{\rm pair} U^{\rm A}_{\rm Nam} \nonumber \\
&=& \Delta_p \left[\cos(\phi_{\rm sc}) \eta_{1} + \sin(\phi_{\rm sc}) \eta_{2} \right] \otimes \left( \begin{array}{cc}
a & b \\
c & d
\end{array} \right),
\end{eqnarray}
where $b$ ($c$) captures the intraband component of the pairing matrix within the conduction (valence) band, and $a$ and $d$ capture the interband components of the pairing. We are solely interested in $b$. After taking $\phi_{\rm sc}+\pi/2 \to \phi_{\rm sc}$ and subsequently setting $\phi_{\rm sc}=0$ without any loss of generality, the reduced BCS Hamiltonian around the Fermi surface on the conduction band reads as 
\begin{equation}
H^{\rm cond}_{\rm BCS}= \xi_{\vec{k}} \eta_3 + \frac{\Delta_p}{k_F} \left( k_x \eta_1 + k_y \eta_2 \right), 
\end{equation}    
where $\xi_{\vec{k}}= v^2 k^2/(2 m) -\tilde{\mu}$ and $k_F=E_k/v$ is the Fermi momentum. This Hamiltonian assumes the form of the topological $p+ip$ paired state, belonging to class D. Next we proceed to perform a similar analysis for the doped class AII Dirac insulator. 

%%%%%%%%%%%%%%%%%%%%%%%%%%%%%%%%%%%%%%%%%%%%%%%%%%%%%%%%%%%%%%%%%%
%%%%%%%%%%%%%%%%%%%%%%%%%%%%%%%%%%%%%%%%%%%%%%%%%%%%%%%%%%%%%%%%%%
%%%%%%%%%%%%%%%%%%%%%%%%%%%%%%%%%%%%%%%%%%%%%%%%%%%%%%%%%%%%%%%%%%
%%%%%%%%%%%%%%%%%%%%%%%%%%%%%%%%%%%%%%%%%%%%%%%%%%%%%%%%%%%%%%%%%%
%%%%%%%%%%%%%%%%%%%%%%%%%%%%%%%%%%%%%%%%%%%%%%%%%%%%%%%%%%%%%%%%%%
\begin{table}
\begin{tabular}{|c|c|c|}
\hline
CF & Pairing Matrix & Near Fermi surface ($\eta_\alpha \otimes \hat{d}^j_{2\times 2}$)  \\
\hline \hline
$\Delta_1^{p}$ &  $\eta_\alpha\Gamma_{00}$ &  $\eta_\alpha \otimes \sigma_0$ \\
$\Delta_2^{p}$ &  $\eta_\alpha\Gamma_{01}$ &  $\eta_\alpha \otimes \sigma_3 (k_x/k_F)$    \\
$\Delta_2^{p}$ &  $\eta_\alpha\Gamma_{32}$ &  $\eta_\alpha \otimes \sigma_3 (k_y/k_F)$   \\
$\Delta_3^{p}$ &  $\eta_\alpha\Gamma_{03}$ &  $m/(k_F v) \; \eta_\alpha \otimes \sigma_0$   \\
$\Delta_4^{p}$ &  $\eta_\alpha\Gamma_{22}$ &  $\eta_\alpha \otimes (k_x \sigma_1 +  k_y \sigma_2)/k_F$  \\
$\Delta_4^{p}$ &  $\eta_\alpha\Gamma_{12}$ &  $\eta_\alpha \otimes (-k_y \sigma_1 +  k_x \sigma_2)/k_F$  \\
\hline
\end{tabular}
\caption{The first column shows the conjugate fields (CFs) associated with six local pairing matrices in a class AII Dirac systems and the second column shows the corresponding matrix operators with $\alpha=1,2$, see Eq.~\eqref{eq:pairQSHINew}. The third column shows the intraband components of the pairing matrices in the conduction band, see Eq.~\eqref{eq:bandprojectionAII}, fostering a Fermi surface. 
}~\label{tab:projectedpairingAII}
\end{table}
%%%%%%%%%%%%%%%%%%%%%%%%%%%%%%%%%%%%%%%%%%%%%%%%%%%%%%%%%%%%%%%%%%
%%%%%%%%%%%%%%%%%%%%%%%%%%%%%%%%%%%%%%%%%%%%%%%%%%%%%%%%%%%%%%%%%%
%%%%%%%%%%%%%%%%%%%%%%%%%%%%%%%%%%%%%%%%%%%%%%%%%%%%%%%%%%%%%%%%%%
%%%%%%%%%%%%%%%%%%%%%%%%%%%%%%%%%%%%%%%%%%%%%%%%%%%%%%%%%%%%%%%%%%
%%%%%%%%%%%%%%%%%%%%%%%%%%%%%%%%%%%%%%%%%%%%%%%%%%%%%%%%%%%%%%%%%%

For this purpose, we rewrite the Nambu-doubled Dirac Hamiltonian in a slightly different form, compared to the one from Eq.~\eqref{HDirNamAII}, as 
\begin{equation}
\hat{h}^{\rm Nam}_{\rm Dir, AII}= v \left( \Gamma_{331} k_x + \Gamma_{302} k_x \right) + (m+b k^2) \Gamma_{303} -\mu \Gamma_{300},
\end{equation}
which we obtain from the one reported in Sec.~\ref{sec:dopedQSHI} after performing a unitary rotation by the unitary matrix $\bar{U}=\eta_0 \otimes (\tau_0 \oplus \tau_1)$. In this basis, the diagonalizing unitary operator reads as $U^{\rm AII}_{\rm Nam}=\eta_0 \otimes \tilde{U}(k_x,k_y,m_{\vec{k}})$, where 
\begin{eqnarray}
&&\tilde{U}(k_x,k_y,m_{\vec{k}}) \\
&=&\left( \begin{array}{cccc}
-\frac{E^-_k}{\sqrt{2 E_k E^-_k}} & 0 & \frac{E^+_k}{\sqrt{2 E_k E^+_k}} & 0 \\
%%%%%%%%%%%%%%%%%%%%%%%%%%%%%%%%%%%%%%%
\frac{v (k_x + i k_y)}{\sqrt{2 E_k E^-_k}} & 0 & \frac{v (k_x + i k_y)}{\sqrt{2 E_k E^+_k}} & 0 \\
%%%%%%%%%%%%%%%%%%%%%%%%%%%%%%%%%%%%%%%
0 & \frac{E^-_k}{\sqrt{2 E_k E^-_k}} & 0 & -\frac{E^+_k}{\sqrt{2 E_k E^+_k}} \\
%%%%%%%%%%%%%%%%%%%%%%%%%%%%%%%%%%%%%%%
0 & \frac{v (k_x-i k_y)}{\sqrt{2 E_k E^-_k}} & 0 & \frac{ v (k_x + i k_y)}{\sqrt{2 E_k E^+_k}} 
\end{array}
\right), \nonumber
\end{eqnarray}
and $E^\tau_k= E_k + \tau (m+b k^2)$ for $\tau=\pm$. After performing a unitary rotation by $U^{\rm AII}_{\rm Nam}$ the Dirac Hamiltonian $\hat{h}^{\rm Nam}_{\rm Dir, AII}$ becomes 
\begin{eqnarray}
&&\left( U^{\rm AII}_{\rm Nam} \right)^\dagger \hat{h}^{\rm Nam}_{\rm Dir, AII} U^{\rm AII}_{\rm Nam} \nonumber \\
&=& \eta_3 \otimes \; \left[ {\rm diag}.(-E_k,-E_k,E_k,E_k)-\mu \Gamma_{00} \right].
\end{eqnarray}
Therefore, the reduced kinetic Hamiltonian describing a collection of gapless excitations around the Fermi surface within the conduction band is given by 
\begin{equation}
\hat{h}^{\rm cond}_{\rm FS, AII}= \left( E_k-\mu \right) \Gamma_{30} \approx \xi_{\vec{k}} \Gamma_{30} + {\mathcal O} (k^4),
\end{equation}
after a large mass expansion of $E_k$, where $\Gamma_{\nu\rho}=\eta_\nu \otimes \sigma_\rho$. The corresponding Nambu-doubled spinor basis takes the form $\Psi^\top_{\rm Nam}=(c_{\vec{k},\uparrow,\rm CB}, c_{\vec{k},\downarrow,\rm CB}, c^\dagger_{-\vec{k},\downarrow,\rm CB}, c^\dagger_{-\vec{k},\uparrow,\rm CB})$, where $c_{\vec{k},\sigma,\rm CB}$ ($c^\dagger_{\vec{k},\sigma,\rm CB}$) is the fermionic annihilation (creation) operator on the CB with momentum $\vec{k}$ and spin projection $\sigma=\uparrow,\downarrow$.

After a unitary rotation by $\bar{U}$ on Eq.~\eqref{eq:pairQSHI}, the effective single-particle Hamiltonian involving all the local pairing now takes the form
\begin{eqnarray}\label{eq:pairQSHINew}
        h^{\rm AII}_{\rm pair} &=& \Delta_1^p \left( \psi^\dagger \Gamma_{\alpha 00}\psi \right) 
				+ \Delta_2^p \; \left[ \left(\psi^\dagger \Gamma_{\alpha 01}\psi \right) + \left( \psi^\dagger \Gamma_{\alpha 32}\psi \right) \right]
				\nonumber \\
				&+& \Delta_3^p \left( \psi^\dagger \Gamma_{\alpha 03}\psi \right)
				+\Delta_4^p \left[ \left( \psi^\dagger \Gamma_{\alpha 22}\psi \right) + \left( \psi^\dagger \Gamma_{\alpha 12}\psi \right) \right],
				\nonumber \\
\end{eqnarray}
where $\psi \equiv \psi_{\rm Nam}$. Any Hermitian matrix describing a local pairing $\hat{h}_{{\rm pair},j}$ with amplitude $\Delta^p_j$ under the unitary rotation by $U^{\rm AII}_{\rm Nam}$ takes the following generic form 
\begin{eqnarray}~\label{eq:bandprojectionAII}
&&\left( U^{\rm AII}_{\rm Nam} \right)^\dagger \hat{h}^{\rm AII}_{{\rm pair},j} U^{\rm AII}_{\rm Nam} \\
&=& \Delta^p_j \left[\cos(\phi_{\rm sc}) \eta_{1} + \sin(\phi_{\rm sc}) \eta_{2} \right] \otimes \left( \begin{array}{cc}
\hat{a}^j_{2\times 2} & \hat{b}^j_{2\times 2} \\
\hat{c}^j_{2\times 2} & \hat{d}^j_{2\times 2}
\end{array} \right). \nonumber
\end{eqnarray}
Here two-dimensional matrices $\hat{d}^j_{2\times 2}$ ($\hat{a}^j_{2\times 2}$) captures the intraband component of the pairing matrices within the conduction (valence) band, and $\hat{b}^j_{2\times 2}$ and $\hat{c}^j_{2\times 2}$ capture their interband components. For $\mu>0$, the Fermi surface is realized within the conduction band. Thus, we are solely interested in $\hat{d}^j_{2\times 2}$. The reduced BCS Hamiltonian associated with each local pairing then takes the form 
\begin{equation}
H^{\rm cond}_{{\rm BCS},j} = \xi_{\vec{k}} \Gamma_{30} 
+ \Delta^p_j \left[\cos(\phi_{\rm sc}) \eta_{1} + \sin(\phi_{\rm sc}) \eta_{2} \right] \otimes \hat{d}^j_{2\times 2}. 
\end{equation}
The intraband components for all the six local pairing matrices are summarized in Table~\ref{tab:projectedpairingAII}. Next we discuss the emergent topology therein.

The two paired states $\Delta^p_1$ and $\Delta^p_3$ are even under parity, spin-singlet (as they appear with the $\sigma_0$ matrix in the spin space), and $s$-wave in nature. Both of them trivially gap out the Fermi surface and thus represent topologically trivial paired state. We note that the gap on the Fermi surface with the $\Delta^p_3$ pairing scales as $m$ (constant Wilson-Dirac mass). Hence, in a Dirac semimetal ($m=0$) this paired state produces a Fermi surface of neutral Majorana fermions.

Each component of $\Delta^p_4$ pairing preserves the rotational symmetry and fully gaps out the Fermi surface for any $\phi_{\rm sc}$. Therefore, in the paired state the reduced BCS Hamiltonian with a convenient choice of $\phi_{\rm sc}=0$ and in a slightly rearranged Nambu-doubled basis given by $\tilde{\Psi}^\top_{\rm Nam}=(c_{\vec{k},\uparrow,\rm CB}, c^\dagger_{-\vec{k},\uparrow,\rm CB},
c_{\vec{k},\downarrow,\rm CB}, c^\dagger_{-\vec{k},\downarrow,\rm CB})$, reads as 
\begin{equation}
H^{\rm cond}_{{\rm BCS},4} = \xi_{\vec{p}} \sigma_0 \otimes \eta_3
+ \frac{\Delta^p_4}{k_F} \; \left[ p_x \sigma_0 \otimes \eta_{1} + p_y \sigma_3 \otimes \eta_{2} \right],
\end{equation}
where $p_x=\cos(\alpha)k_x-\sin(\alpha)k_y$ and $p_y=\cos(\alpha)k_x+\sin(\alpha)k_y$ with $\alpha$ as the internal angle between two components of $\Delta^p_4$, which gets locked spontaneously. In this basis the paired state assumes the form of the topological $p \pm ip$ pairing, preserving the time-reversal symmetry and belonging to class DIII. The pairing symmetry is $p+ip$ and $p-ip$ for spin projections $\sigma=\uparrow$ and $\downarrow$, respectively.

Each component of $\Delta^p_2$ pairing breaks the rotational symmetry (thus representing a nematic pairing) and gives rise to point nodes on the Fermi surface. Nevertheless, this paired state can fully gap out the Fermi surface when the difference in the superconducting phase between its two components is $\pi/2$, indicating the breakdown of the time-reversal symmetry. In a newly defined Nambu-doubled basis $\tilde{\Psi}^\top_{\rm Nam}=(c_{\vec{k},\uparrow,\rm CB}, c^\dagger_{-\vec{k},\uparrow,\rm CB}, c_{\vec{k},\downarrow,\rm CB}, -c^\dagger_{-\vec{k},\downarrow,\rm CB})$ the reduced BCS Hamiltonian then reads as 
\begin{equation}
H^{\rm cond}_{{\rm BCS},2} = \sigma_0 \otimes \left\{ \xi_{\vec{p}}  \eta_3
+ \frac{\Delta^p_2}{k_F} \; \left[\cos(\alpha) p_x \eta_{1} + \sin(\alpha) p_y \eta_{2} \right] \right\},
\end{equation}
where $\alpha$ ($\neq \pm \pi/4$ in general) is the internal angle, manifesting the lack of rotational symmetry in the paired state. The paired state corresponds to two identical copies of the $p+ip$ pairing and its topological properties are identical to those for the $d+id$ superconductor (for any $\alpha$). This paired state belongs to class C and the requisite SU(2) symmetry is generated by ${\boldsymbol \sigma} \otimes \eta_0$~\cite{TITSC:13}.

As a final remark, we note that throughout this Appendix we assumed that the underlying Fermi surface is simply connected and established the topological nature of various local pairings, which are then endowed by nontrivial topological invariant. However, when the Fermi surface becomes annular, the net topological invariant of all the paired states becomes zero. This observation can be appreciated qualitatively in the following way. Notice that the curvature of the annular Fermi surface is opposite at the inner and outer Fermi ring. Consequently, their contributions to the topological invariant cancel each other, yielding a trivial $p$-wave superconducting states. A more detailed discussion on this issue is left for a separate investigation.

\begin{widetext}

\section{Contributions from one-loop Feynman diagrams}~\label{append:Feynman}

In this Appendix, we display the contributions from the one-loop Feynman diagrams shown in Figs.~\ref{Fig:FeynDiag_Interaction} and~\ref{Fig:FeynDiag_Susceptibility}. For this purpose, we consider two local four-fermions interactions, given by $g_{_M} (\Psi^{\dagger} M \Psi)^2$ and $g_{_N} (\Psi^{\dagger} N \Psi)^2$, where $g_{_{M}}$ and $g_{_{N}}$ are the corresponding coupling constants, respectively. Throughout this Appendix, we assume that the fermion spinors $\Psi$ and $\Psi^\dagger$ are defined in the Nambu-doubled basis, $M$ and $N$ are Hermitian matrices that are also expressed in the Nambu-doubled basis. The corresponding fermionic Green's function in the Nambu doubled basis is denoted by $G(i \omega_n,\vec{k})$, where $\omega_n$ are the fermionic Matsubara frequencies. We do not display their forms explicitly in this Appendix, which can be readily obtained in Sec.~\ref{sec:dopedQAHI} for class A and Sec.~\ref{sec:dopedQSHI} for class AII systems. Then the contributions from the Feynman diagram from Figs.~\ref{Fig:FeynDiag_Interaction}(b)-\ref{Fig:FeynDiag_Interaction}(f) are schematically given by 
\allowdisplaybreaks[4]
\begin{align}
(3b)& = \Psi^{\dagger} \left[  g_{_M} \; N_{\rm comb}^{\rm Self-energy} \; \sum_{n=-\infty}^{\infty}\int_{\Lambda e^{-l}}^{\Lambda}\frac{k dk}{2 \pi} \int_0^{2 \pi} \frac{d \phi_{\vec{k}}}{2 \pi} \:\:\: M  G(i \omega_n, \vec{k}) M \right] \Psi, \\
(3c)&= \left[-\frac{N_{\rm comb}^{\rm Bubble}}{2} \; g_{_M} g_{_N} \; \; \frac{1}{2} \; \sum_{n=-\infty}^{\infty}\int_{\Lambda e^{-l}}^{\Lambda}\frac{k dk}{2 \pi} \int_0^{2 \pi} \frac{d \phi_{\vec{k}}}{2 \pi} \:\:\: \textrm{Tr}[M  G(i \omega_n, \vec{k}) N G(i \omega_n, \vec{k}) ] \right] (\Psi^{\dagger} M \Psi) (\Psi^{\dagger} N \Psi), \\
(3d)&= \left[ \frac{N_{\rm comb}^{\rm Vertex}}{2} g_{_M} g_{_N} \; (2-\delta_{M,N}) \; \sum_{n=-\infty}^{\infty}\int_{\Lambda e^{-l}}^{\Lambda}\frac{k dk}{2 \pi} \int_0^{2 \pi} \frac{d \phi_{\vec{k}}}{2 \pi} \:\:\: (\Psi^{\dagger} [N  G(i \omega_n, \vec{k}) M G(i \omega_n, \vec{k}) N ] \Psi) \right] \; (\Psi^{\dagger} M \Psi), \\
(3e)&= \frac{N_{\rm comb}^{\rm Ladder}}{2} g_{_M} g_{_N} \; (2-\delta_{M,N}) \; \sum_{n=-\infty}^{\infty}\int_{\Lambda e^{-l}}^{\Lambda}\frac{k dk}{2 \pi} \int_0^{2 \pi} \frac{d \phi_{\vec{k}}}{2 \pi} \:\:\:  (\Psi^{\dagger}[N  G(i \omega_n, \vec{k}) M  ] \Psi) \: ( \Psi^{\dagger} [ M G(i \omega_n, \vec{k})N ] \Psi), \\
\text{and} & \;\; (3f) =\frac{N_{\rm comb}^{\rm Crossing}}{2} g_{_M} g_{_N} \; (2-\delta_{M,N}) \; \sum_{n=-\infty}^{\infty}\int_{\Lambda e^{-l}}^{\Lambda}\frac{k dk}{2 \pi} \int_0^{2 \pi} \frac{d \phi_{\vec{k}}}{2 \pi} \:\:\: (\Psi^{\dagger}[N  G(i \omega_n, \vec{k}) M  ] \Psi) (\Psi^{\dagger} [ N G(-i \omega_n, -\vec{k})M ] \Psi), 
\end{align}
respectively, where $\phi_{\vec{k}}=\tan^{-1}(k_y/k_x)$. In the above expressions $\Psi$ and $\Psi^\dagger$ are the \emph{slow} fermionic fields with momentum $|\vec{k}|<\Lambda e^{-\ell}$ and the fermionic Green's functions are obtained by contracting the \emph{fast} fields with $\Lambda e^{-\ell}<|\vec{k}|<\Lambda$. The factor of $1/2$ in the last four equation arises from the Taylor expansion of $S_{\rm int}$ to the quadratic order, the extra ``minus" sign in the second equation stems from the fermion bubble where the additional factor of $1/2$ cancels the contribution from the Nambu doubling. This diagram yields nontrivial contribution only when $M =N$. Then the combinatorial factor of the Feynman diagrams are given by $N_{\rm comb}^{\rm Self-energy}=2$, $N_{\rm comb}^{\rm Bubble}=4$, $N_{\rm comb}^{\rm Vertex}=8$ ($4$) for $M=N$ ($M \neq N$), and $N_{\rm comb}^{\rm Ladder}=N_{\rm comb}^{\rm Crossing}=4$ ($2$) for $M=N$ ($M \neq N$).

To arrive at the contributions from the Feynman diagrams shown in Fig.~\ref{Fig:FeynDiag_Susceptibility}, here we consider one source field $\Delta_{_{\mathcal O}} (\Psi^{\dagger} {\mathcal O} \Psi)$, where ${\mathcal O}$ is yet another Hermitian matrix represented in the Nambu-doubled basis and $\Delta_{_{\mathcal O}}$ is the corresponding conjugate field. Then contributions from Figs.~\ref{Fig:FeynDiag_Susceptibility}(b) and\ref{Fig:FeynDiag_Susceptibility}(c) are respectively given by
\allowdisplaybreaks[4]
\begin{align}
(4b)&=\left[ -2 \; \frac{N_{\rm comb,source}^{\rm Bubble}}{2} g_{_M} \; \frac{1}{2} \; \sum_{n=-\infty}^{\infty}\int_{\Lambda e^{-l}}^{\Lambda}\frac{k dk}{2 \pi} \int_0^{2 \pi} \frac{d \phi_{\vec{k}}}{2 \pi} \:\:\:  \textrm{Tr}[M G(i \omega_n, \vec{k}) {\mathcal O} G(i \omega_n, \vec{k})] \right] \:\: \left( \Psi^{\dagger} M \Psi \right) \\
\text{and} & \:\: (4c)=2\; \frac{N_{\rm comb,source}^{\rm Vertex}}{2} g_{_M} \; \sum_{n=-\infty}^{\infty}\int_{\Lambda e^{-l}}^{\Lambda}\frac{k dk}{2 \pi} \int_0^{2 \pi} \frac{d \phi_{\vec{k}}}{2 \pi} \:\:\: (\Psi^{\dagger}[M  G(i \omega_n, \vec{k}) {\mathcal O} G(i \omega_n, \vec{k})M ]\Psi),
\end{align}
where $N_{\rm comb,source}^{\rm Bubble}=N_{\rm comb,source}^{\rm Vertex}=2$, the extra factor of $2$ in the numerator arises from the Taylor expansion, and the factor of $1/2$ in the first equation cancels the Nambu doubling.

\end{widetext}

%%%%%%%%%%%%%%%%%%%%%%%%%%%%%%%%%%%%%%%%%%%%%%%%%%%%%%%%%%%%%%%%%%%%%%%%%%%%%%%%%
%%%%%%%%%%%%%%%%%%%%%%%%%%%%%%%%%%%%%%%%%%%%%%%%%%%%%%%%%%%%%%%%%%%%%%%%%%%%%%%%%
%%%%%%%%%%%%%%%%%%%%%%%%%%%%%%%%%%%%%%%%%%%%%%%%%%%%%%%%%%%%%%%%%%%%%%%%%%%%%%%%%

\end{document}